\documentclass[a4paper,fleqn,usenatbib]{mnras}

\usepackage[T1]{fontenc}
\usepackage{ae,aecompl}

\usepackage{graphicx}	
\usepackage{amsmath}	
\usepackage{amssymb}	

\usepackage{ulem}
\usepackage{multicol}

\newcommand{\be}{\begin{equation}}
\newcommand{\ee}{\end{equation}}

\let\vec\mathbf

\title[The effect of fluctuating fuzzy axion haloes]{The effect of fluctuating fuzzy axion haloes on stellar dynamics: a stochastic model}

\author[Amr El-Zant, Jonathan Freundlich, Fran\c coise Combes, Anaelle Halle]{ 
 Amr A. El-Zant$^{1}$\thanks{E-mail: amr.elzant@bue.edu.eg}, Jonathan Freundlich$^{2,3}$, Fran\c coise Combes$^{2,4}$, Anaelle Halle$^{2,4}$
\\
$^{1}$ Centre for Theoretical Physics, The British University in Egypt, Sherouk City 11837, Cairo, Egypt\\
$^{2}$ LERMA, Observatoire de Paris, CNRS, Sorbonne Universit\'es, UPMC Univ. Paris 06, F-75014, Paris, France\\
$^{3}$ Centre for Astrophysics and Planetary Science, Racah Institute of Physics, The Hebrew University, Jerusalem 91904, Israel\\
$^{4}$ Coll\`ege de France, PSL Research University, F-75005, Paris, France\\
}

\date{Accepted XXX. Received YYY; in original form ZZZ}

\pubyear{2018}

\begin{document}
\label{firstpage}
\pagerange{\pageref{firstpage}--\pageref{lastpage}}

\maketitle

\begin{abstract}
Fuzzy dark matter of ultra-light axions has gained  attention, 
largely in light of the galactic scale problems associated with cold dark matter.  
But the large de Broglie wavelength, believed to possibly alleviate these problems, 
also leads to fluctuations that place constraints on ultra-light axions. 
We adapt and extend a method, previously devised to describe 
the effect of gaseous fluctuations on cold dark matter cusps,  in order to determine
the imprints of ultra-light axion haloes on  the motion of classical test particles.
We first evaluate the effect of  fluctuations in a statistically homogeneous medium of classical particles,  
then in  a similar system of ultra light axions.  
In the first case, one recovers the classical two body relaxation time (and diffusion coefficients) from white noise
density fluctuations. In the second  situation, the fluctuations are not born of discreteness noise but from the finite de Broglie wavelength; 
correlation therefore exists over this scale,  while white noise is retained on larger scales, elucidating the correspondence with 
classical relaxation.  The resulting density power spectra and correlation functions  are compared with those inferred from numerical simulations, and 
the relaxation time arising from the associated  potential fluctuations is  evaluated. 
We then apply our results to estimate the heating of disks embedded in axion dark haloes. We 
find that this implies an axion mass  $m \ga 2 \times 10^{-22} {\rm eV}$.  
We finally apply our model to the case of the central cluster of Eridanus II,   
confirming that far stronger constraints on $m$ may in principle be obtained, and discussing the 
limitations associated with the assumptions leading to these. 
\end{abstract}

\begin{keywords}
dark matter -- galaxies: haloes -- galaxies: kinematics and dynamics --  galaxies: evolution -- galaxies: formation 
\end{keywords}





\section{Introduction}

The cold dark matter based scenario has developed into a highly successful model of structure 
formation (\citealp{White_F2012}).  
The weakly interacting massive particles (WIMPs) at the core of this scenario can be produced with the right
abundance, and a cross section of the order expected of standard weak interaction, 
from an early thermal equilibrium in the radiation era. Yet, extensive direct detection experiments and collider searches 
have significantly constrained the expected (mass and cross section) parameter space for such particles 
\citep{Wimp2018, DM_Coll2018, Waning_Wimp2018}. 
In addition, from the astrophysical viewpoint, WIMP-based virialized structures suffer from several 
'small scale' problems; such as the cusp-core problem, the 'too big to fail' problem and the overabundance 
of subhaloes (\citealp{delPopolo2017}; \citealp{Bullock_B2017}, for recent reviews).  

There are proposed solutions for such (very possibly related)
problems in terms of baryonic physics. For example  through dynamical friction mediated coupling 
with baryonic clumps
\citep{Zant2001, Zant2004, Tonini2006, RomanoDiaz2008, Goerdt2010, Cole2011, delPopolo2014, Nipoti2015}, 
or through gas fluctuations arising from star formation or active galactic nuclei~\citep{Read2005, Mashchenko2006, Mashchenko2008, Peirani2008, Pontzen2012, Governato2012, Zolotov2012, Martizzi2013, Teyssier2013, Pontzen2014, Madau2014, Freundlich2019}.
Recently,~\cite{EZFC} developed a stochastic model for such fluctuations and their effect on the cold dark matter halo 
cusp, in an attempt to understand the mechanism of central halo 'heating' and core formation 
from first principles.  In the present study, we find that it has wider applications, particularly concerning 
potentially observable dynamical effects of the fuzzy dark matter (FDM) composed of ultralight axions.

Ultra light axions,  with boson mass $\sim 10^{-22} {\rm eV}$, have long been considered as dark matter candidates
 in connection to the problems facing CDM
(e.g., \citealp{Peebles2000, Goodman2000, Hu2000,  Schive2014, Marsh2014,  Hui_etal2017,
Chavanis2018, BaldiLyman, Mocz2019}. 
The earlier history of the subject is reviewed in~\citealp{Chavanis2011} 
and~\citealp{Lee2018}, the former also considering the case of systems with short range interactions).
The long de Broglie wavelength associated with their tiny mass results in 'fuzziness' in their position 
that implies finite density Bose-condensate halo cores, instead of cusps, and the dissolution of smaller subhaloes.  
The field associated with such axions may also  play a role in sourcing inflation or late dark energy, and non-thermal production 
implies that they can come with the right abundance and dynamically behave as WIMPs on larger scales despite their small mass~\citep{Marsh2016, Marsh2017}. 
In light of the relative waning of the WIMP paradigm, light axion FDM has recently gained ground as viable contenders, 
namely as an alternative to the baryonic solutions to galactic scale problems from 
within the CDM scenario. 
In this context, they are one of several  particle physics based proposals,  including  
warm dark matter \citep[e.g., ][]{Colin2000, Bode2001, Schneider2012, Maccio2012b, Shao2013, Lovell2014, El-Zant2015} and
self-interacting dark matter  \citep[e.g.,][]{Spergel2000,Burkert2000,Kochanek2000,Miralda2002, Peter2013, Zavala2013, Elbert2015}.

Nevertheless, although the  large de Broglie wavelength associated with the ultra-light halo bosons 
can in principle help solve the core/cusp and potentially associated small scale problems, it is not clear if the
scaling of core radii with mass inferred from (dark matter only) simulations
can be made to agree with observed scalings (\citealp{BarnBH19,Hertz2018,MohSperg19,Robles2019}), and there are additional constraints regarding the particle mass
arising from Lyman-$\alpha$ and 21 cm observations  (e.g., 
\citealp{Lyman1_2017};  \citealp{21cm1_2018}; \citealp{Hui_21cm}) and recently 
from  environment around  supermassive black holes (e.g., \citealp{Superad}; \citealp{BarnBH19}; \citealp{NusserBH19}; \citealp{EllioMocz19}). 
 
A large de Broglie wavelength also leads to reduction of bound substructure,  in apparent agreement with observations. 
But  as bound substructure is replaced by broad interference patterns \citep[e.g., ][]{Schive2014I}, this is accompanied by density fluctuations that may lead to observable effects on the baryonic components of galaxies and place constraints on the masses  of FDM particles (\citealp{Hui_etal2017}; \citealp{BOFT}; \citealp{AmLoeb2018}; \citealp{Marsh2018}~\citealp{Church_2018}).  This leads to the following conundrum. 
 For FDM to be effective in solving the small scale 
problems of CDM,  the de Broglie wavelength must be of the order of the scales at which the problems appear (i.e., kpc scale); the masses of the axions must therefore be small enough for their wavelength to reach such scales. 
But then the associated fluctuations are large, and a delicate balance seems required in order 
to solve the galactic scale problems of CDM and at the same time not overproduce fluctuations (and avoid unobserved consequences on the baryonic components of galaxies).  In other words, sufficiently suppressing such fluctuations may imply axion masses that are too large to solve the small scale problems of the WIMP based structure formation scenario.

The  density fluctuations give rise to potential fluctuations, much in the same manner as 
the gaseous fluctuations studied in~\cite{EZFC}.  Here, we make use of this fact in order to 
study their effect on the stellar dynamics of galaxies with the aforementioned problem in mind.  
For this purpose, we adopt and extend the methods outlined there
to physically quite distinct, but formally related, contexts: 
first to derive the standard two body relaxation time, by assuming delta correlated density fluctuations (Section~\ref{sec:white});  
then to estimate the density and force fluctuations in FDM haloes, and to calculate the  
associated correlation functions and relaxation time of a classical test particle subject to FDM halo 
fluctuations, pointing out differences and similarities with classical two body relaxation due to discreteness (white) noise (Section~\ref{sec:axion_fluc}); 
and  finally, as an application, to estimate the effect of fluctuations of an 
FDM halo on the Milky Way disk in light of recent observations of the stellar velocity dispersion, putting constraints
on the minimal mass of plausible FDM particles (Section~\ref{sec:disk}).  In Section~\ref{sec:comparison} we compare 
the resulting constraints to those from other work. In that section, we also discuss the predictions of our present model 
regarding the expansion of the central star cluster of the dwarf galaxy Eridanus II, as studied in~\cite{Marsh2018}
by adopting the original formulation of~\cite{EZFC}. The technical details associated with that discussion are given in 
Appendix~\ref{App:Eri}. Section~\ref{sec:conc} summarizes our results and outlines related conclusions.

\section{White noise and two body relaxation}
\label{sec:white}

We start by considering how the basic theoretical setup introduced in~\cite{EZFC}
can be directly applied  to derive the standard two body 
relaxation time for the case of a test star moving through an infinite system of randomly distributed 
'field stars' (point particles), of average spatial mass density $\rho_0$. 

As in the aforementioned work, we Fourier expand the 
potential $\Phi$ and density contrast $\delta = \frac{\rho({\bf r})}{\rho_0} -1$, so that
\begin{equation}
\Phi  ({\bf r}, t) = \frac{V}{(2 \pi)^3}  \int \Phi_{\bf k} (t) e^{i {\bf k . r}} d {\bf k},
\label{eq:phirk}
\end{equation} 
and 
\begin{equation}
\delta ({\bf r}, t) = \frac{V}{(2 \pi)^3} \int \delta_{\bf k} (t) e^{i {\bf k . r}} d {\bf k}.
\label{eq:rhork}
\end{equation} 
Here the volume $V$, previously taken to be much larger than the largest fluctuation scales, should be 
understood to be arbitrarily large when white noise is considered (as we will see below, the largest 
relevant fluctuation scales are then effectively determined by the argument of a Coulomb logarithm).  

We note at the outset that this formulation already incorporates a form of the 'Jean's swindle', whereby 
the potential $\Phi$ is considered to be solely due to fluctuations around the average mass density in an infinite 
medium that tends to homogeneity on larger scales (e.g., Binney \& Tremaine 2008; \citealp{BOFT}).   
However, as opposed to the standard Jeans swindle, it is not used to calculate the self-gravity 
of the fluctuations but their gravitational effects on a test particle, while neglecting their own self-gravity.  
It therefore incorporates the analogous 'Chandrasekhar swindle', implicit 
in the derivation of the standard two body relaxation time, which evaluates the effect of finite-$N$ fluctuations 
in an infinite statistically homogeneous medium with no mean field contribution. 
As all realistic self-gravitating systems are inhomogeneous,  application of the 'Jeans-Chandrasekhar' 
swindle implicitly involves a local assumption. A systematic study of the 
steps leading to this approximation in the context of kinetic theory, as well as an exposition of the rich history 
of treating fluctuations and associated relaxation in gravitating systems,  can be found in~\cite{Chavanis2013}.  
Here, we just further note that it is in principle possible to incorporate collective effects, due to self gravity,
into the kinetic theory of gravitating systems --- either while maintaining the assumption of large scale homogeneity (\citealp{Weinberg1993}),
or for inhomogeneous systems with integrable mean field dynamics, where action angle variables can be used
(\citealp{Heyvaerts2010, Chavanis2012}) --- though the resulting formulations  require  
considerable additional mathematical sophistication and are therefore more difficult to extract information from in practice.

In~\cite{EZFC} the $\Phi$ and $\delta$ were initially 
assumed to be time independent. The time dependence was then introduced through 
the sweeping ansatz used and tested in turbulence theory, whereby small scale fluctuations 
are 'swept' --- that is passively advected ---  by large scale velocity fields, which can either
represent mean bulk flows~\citep{Taylor1938}, or random large scale flows characterized by a velocity 
dispersion \citep{Kraichnan1964, Tennekes1975}. 
The effective assumption in~\cite{EZFC} was 
that all modes are transported by the same sweeping speed.  The model is however
easy to generalize to the case when each mode travels with its own  
velocity, drawn from a given distribution~\citep[e.g., ][]{Wilczek2012, Wilczek2014}.
Here we generalize this sweeping hypothesis to the transport of density fluctuations, 
focusing on classical point particles and FDM systems.

We define a two time density fluctuation power spectrum that is given through
the ensemble average 
\begin{equation}
\mathcal{P}({\bf  k}, t, t^{\prime})   = V  \langle  \delta_{\bf k} (t)   \delta^{*}_{\bf k} (t^{\prime})    \rangle.
\end{equation}
For a stationary stochastic process, the time dependence  
manifests itself solely in terms of $t - t^{\prime}$, for all $t^{\prime}$, reflecting 
homogeneity in time.  We assume that this is the case here, setting $t^{\prime} = 0$ without loss of generality.
In particular, the equal time power spectrum is
\begin{equation}
\mathcal{P}({\bf  k}, 0)   = V  \langle  |\delta_{\bf k} (0)|^2    \rangle.
\end{equation}

The  components $\Phi_{\bf k} (t)$ and $\delta_{\bf k} (t)$ are related via the Poisson equation
$ 
\nabla^2 \Phi= 4 \pi G \rho_0 \delta, 
$ 
through
\begin{equation}
\Phi_{\bf k} (t)  = -4 \pi G \rho_0 \delta_{\bf k} (t) k^{-2},
\label{eq:phikk}
\end{equation}
where $ k  = |{\bf k}|$.
For a configuration  that is isotropic on large scales, 
the force power spectrum is related to the potential fluctuations by 
\begin{equation}
\label{eq:pfk}
\mathcal{P}_F ({ k} , t)  = V { k}^2 \langle \Phi_{ k} (0) \Phi_{k} (t)  \rangle = (4 \pi G \rho_0)^2 k^{-2} \mathcal{P}({\bf  k}, t).
\end{equation}
For a system that is homogeneous on large scales, the force correlation function, which is
the inverse Fourier transform of the force power spectrum, is given by
\begin{align}
\nonumber
\langle {\bf F} (0, 0) . {\bf F} (r, t)\rangle & = 
\frac{1}{(2 \pi)^3} \int  \mathcal{P}_F ({ k} , t)  e^{i {\bf k}. {\bf r}} d {\bf k}\\
& =   \frac{1}{(2 \pi)^3} \int  \mathcal{P}_F ({ k} , t)  \frac{\sin (k r)}{k r} 4 \pi k^2 d k. 
\label{eq:FCF}
\end{align}

\subsection{Fixed velocities} 
\label{sec:fixed}

We first consider the simplest case,  in which all modes move with the same velocity. 
This connects the sought after derivation of the two body relaxation time to the 
situation  studied in~\cite{EZFC}, where we assumed a spatial (one time) power law spectrum of the form 
$\langle  |\delta_k (0)|^2  \rangle = C k^{-n}$, and introduced 
the time dependence through a constant speed sweeping hypothesis.  We proceed in an 
analogous manner here, but focus on the special case of a white noise density spectrum ($n = 0$), 
appropriate  of the expected spectrum of randomly scattered  
point masses that we take to represent the 'field stars' through which 
a test particle moves. As in the case of the standard derivation of the 
two body relaxation time, the system of field particle point masses 
is assumed to be spatially homogeneous, beyond the Poisson noise, and the 
test particle's unperturbed motion is rectilinear with constant 
velocity ${\bf v}_p$. 

In this case, and introducing maximum and minimal 
cutoff scales, the spatial force correlation function can be written as  
\begin{align}
\nonumber
\langle {\bf F}(0, 0) . {\bf F}(r, 0) \rangle =   \frac{D}{r} \int_{k_m}^{k_x }       \frac{\sin (k r)}{k} d k\\ 
=    \frac{D}{r}  \left[{\rm Si} (k_x r) - {\rm Si} (k_m r) \right] 
\end{align}
where $D = 8 (G \rho_0)^2  \mathcal{P}({\bf  k}, 0)$ and ${\rm Si}$ refers to the sine integral function. For a white noise power spectrum, $\mathcal{P}({\bf  k}, 0)$ is constant.
The force correlation function can in principle   be inserted into the stochastic equation 
\citep[e.g., ][]{Osterbrock1952,EZFC}
\begin{equation}
\langle (\Delta v_p)^2 \rangle = 2 \int_0^T (T - t)  \langle {\bf F}(0) . {\bf F}(t) \rangle d t, 
\label{eq:corrsp}
\end{equation}
in order to obtain the velocity variance that the test particle 
acquires as a result of its motion through 
fluctuating potential of the randomly distributed field particles.  

To do this,  one 
has to transform the spatial correlation function into one involving time. 
In this regard, it
is important to note  that ${\bf F}(t)$ refers to the force at time $t$ 
on a test particle, and is thus evaluated 
along a particle 
trajectory.  As the field is time dependent and the test particle also moves, 
in fixed (Eulerian) coordinates the relevant force entering into 
equation(\ref{eq:corrsp}) is 
${\bf F} = {\bf F} ({\bf r}_p, t)$, where  
${\bf r}_p (t)$ refers to the position of the particle at time $t$. This is assumed to be along 
the unperturbed trajectory  (a straight line).

In~\cite{EZFC} we incorporated both the time dependence due to the motion of the test particle and 
the evolution of the fluctuating field by assuming that that latter is 'swept', moving 
with its statistical properties 'frozen in', 
with constant speed $v_r$
relative to the test particle. It is then a simple matter to relate $\langle {\bf F}(0, 0) . {\bf F}(r, 0) \rangle$
and $\langle {\bf F}(0) . {\bf F}(r_p, t) \rangle$.
In the current context, an analogous assumption is that the field stars have negligible 
velocity dispersion and common 'bulk' velocity ${\bf v}_f  = {\bf v}_r  + {\bf v}_p$, with ${\bf v}_r$
being the relative velocity.~\footnote{In what follows 
we drop the subscript $f$ in referring to the field particle velocities while the test particle velocity will  still 
be denoted by ${\bf v}_p$.}
In this case, along a  particle trajectory, the force correlation function is 
\begin{align}
\nonumber
\langle {\bf F}(0)~.~{\bf F}(r = v_r t) \rangle & =    \frac{D}{v_r t}     \int_{k_m}^{k_x }       \frac{\sin (k v_r t)}{k} d k\\ 
& =     \frac{D}{v_r t}  \left[{\rm Si} (k_x v_r t) - {\rm Si} (k_m v_r t) \right].
\end{align}
This may then be inserted into (\ref{eq:corrsp}), to obtain the velocity dispersion 
that the test particle acquires as a result of the  
application of the stochastic force described by the described correlation function:
\begin{equation}
\langle (\Delta v_p)^2 \rangle = 2 D \int \frac{(T - t)}{v_r t} \left[{\rm Si} (k_x v_r t) - {\rm Si} (k_m v_r t) \right] d t.
\label{eq:detdisp}
\end{equation}
As the Sine integral functions  in this equation converge to $\pi/2$ when $k_x v_r t, k_m v_r t \gg 1$, 
in this diffusion limit the velocity dispersion increase is dominated by the non-transient term 
(involving $T$ rather than $-t$ in the bracket multiplying the correlation function in \ref{eq:detdisp}).
As detailed in Appendix~\ref{app:2body}, one then finds that
\begin{equation}
\langle (\Delta v_p)^2 \rangle = \frac{\pi  D}{v_r} T  \ln \frac{k_x}{k_m}, 
\label{eq:tbs}
\end{equation}
which has the form of the standard two body relaxation time if the maximal and minimal 
fluctuation scales are identified with the maximal and minimal impact parameters 
of classical theory.  

For a system of field point particles of mass $m$, 
randomly distributed with uncorrelated positions and average homogeneous density $\rho_0$, 
the (equal time) spatial density correlation 
function is 
\begin{equation} 
\langle \rho (0,0) \rho (\vec{r}, 0) \rangle =
\rho_0^2~\langle \delta (0,0) \delta (\vec{r}, 0) \rangle = m \rho_0 \delta_D (\vec{r}), 
\label{eq:Decorr}
\end{equation}
where $\delta_D$ refers to the Dirac delta function.  
The associated power spectrum is  simply 
\begin{equation} 
\mathcal{P}({k}, 0) = \frac{m}{\rho_0}. 
\end{equation}
This implies that the density fluctuations of the white noise 
are equally distributed among the modes such that 
$\langle |\delta|^2 \rangle = 1/N$, with $N=\rho_0 V / m$ the number of particles within the volume $V$, and $D = 8 G^2 m \rho_0$. 
In this case 
\begin{equation}
\langle (\Delta v_p)^2 \rangle = \frac{8 \pi G^2 \rho_0 m}{v_r}   T \ln \Lambda,  
\label{eq:tbs2}
\end{equation}
where we have set $\Lambda = \frac{k_x}{k_m} = \frac{\lambda_{\rm max}}{\lambda_{\rm min}}$, 
$\lambda_{\rm max}$ and $\lambda_{\rm min}$ being the maximal and minimal fluctuation wavelengths.
Assuming that the test particle speed $\sim v_r \sim v$, then the timescale 
for the RMS perturbation to the velocity described by (\ref{eq:tbs}) to reach a  
typical speed  $v$ is 
\begin{equation}
t_r = \frac{v^3}{8 \pi G^2 \rho_0 m  \ln \Lambda},
\label{eq:tbr}
\end{equation}
which is the standard formula for the two body relaxation time.

\subsection{Distribution of field particle velocities} 
\label{sec:dist}

So far we have assumed that all modes are 'swept' by the same 
constant velocity field,  and by implication (in the case of white noise density 
fluctuations) that all field particles 
had the same velocity. It is possible however to extend the sweeping picture 
to the case when each mode has its own 'advection' velocity. In the current context 
this  translates into extending the formulation to include a velocity 
distribution for the field particles. As each mode is passively advected with velocity ${\bf v}$, mass conservation requires
that
\begin{equation}
i \frac{\partial \rho_{\bf k} (t)}{\partial t} =  {\bf k}.{\bf v}~\rho_{\bf k} (t),
\end{equation}
where  $\rho_{\bf k} (t) = \rho_0 \delta_{\bf k} (t)$. This has for solution
\begin{equation}
\rho_{\bf k} (t) =\rho_{\bf k} (0)  e^{- i {\bf k}.{\bf v}t}.
\label{eq:sweeprho}
\end{equation}
Thus 
\begin{equation}
\langle \rho_{\bf k} (0) \rho^*_{{\bf k ^{\prime}}} (t) \rangle 
= \rho_0^2 \langle  \delta_{\bf k} (0) \delta^*_{{\bf k}^{\prime}} (t) \rangle
 = \rho_0^2 \langle  e^{i {\bf k^{\prime}} .{\bf v} t} \delta_{\bf k} (0) \delta^*_{{\bf k}^{\prime}} (0) \rangle. 
\label{eq:sweepcorr}
\end{equation}
For a homogeneous (in both space and time) stochastic process defined over an infinite spatial volume, 
equation~(\ref{eq:rhork}) implies that
\begin{equation}
V^2 \langle  \delta_{\bf k} (0) \delta^*_{{\bf k}^{\prime}} (t) \rangle
= (2 \pi)^3 \mathcal{P}({\bf  k}, t) \delta_D ({\bf k} - {\bf k}^{\prime}),
\label{eq:PSdelt}
\end{equation}
as the power spectrum is the Fourier transform of the spatial correlation function.

The density correlation function takes a particularly simple form if the density fluctuations are (at least initially) 
uncorrelated with the velocities; that is, each mode can have any velocity.   
This will be the case if there is no explicit dispersion relation tying
${\bf v}$ and ${\bf k}$, which is the case of classical particle systems, but not 
when the fluctuations have  quantum origin as discussed later. When no such correlation exists, one can
use equations~(\ref{eq:sweepcorr}) and~(\ref{eq:PSdelt})  to derive
\begin{equation}
\mathcal{P} ({\bf  k}, t) = \mathcal{P}({\bf  k}) \langle e^{- i {\bf k}.{\bf v}~t} \rangle;  
\label{eq:simp}
\end{equation}
where, again assuming a homogeneous process, we used 
$V^2 \langle \delta_{\bf k} (0)  \delta^*_{\bf k^{\prime}} (0) \rangle =(2 \pi)^3 \mathcal{P}({\bf  k}) \delta_D ({\bf k} - {\bf k}^{\prime})$
and integrated over the delta functions.

Assuming a mass normalized velocity distribution function $f ({\bf v})$, such that $\int f ({\bf v}) d {\bf v} = \rho_0$, this leads to 
\begin{equation}
\mathcal{P} ({\bf  k}, t) = \frac{\mathcal{P}({\bf  k})}{\rho_0} \int e^{- i {\bf k}.{\bf v}~t} f ({\bf {v}}) d {\bf v}.
\end{equation}
For the case of delta correlated white noise (of equation~\ref{eq:Decorr})
\begin{equation}
\mathcal{P}({\bf  k}, t) =
\frac{m}{\rho_0^2} ~\int e^{- i {\bf k}.{\bf v}~t} f ({\bf {v}}) d {\bf v}.
\label{eq:2b_ps}
\end{equation}
Note that this leads to the same wavenumber-frequency power spectrum $\hat{C} ({\bf k}, \omega)$ 
as given by equation (23) of~\cite{BOFT}. 
Using equations~(\ref{eq:pfk}) and (\ref{eq:FCF})
one can obtain the force correlation function
\begin{equation}
\label{eq:F00Frt}
\langle {\bf F} (0, 0) . {\bf F} (r, t)\rangle =
8~G^2 m  \int  f ({\bf v}) d {\bf v} \int \frac{\sin (k |{\bf r}/t -  {\bf v}| t)}{k |{\bf r}/t -  {\bf v}| t} d k, 
\end{equation}
which assumes an isotropic medium (but not necessarily isotropic velocities)~\footnote{Integration over 
$k$ ($[0, \infty]$) shows this to be equivalent to the form found by~\cite{Cohen}, who directly 
evaluated correlations of the Newtonian force (his equation 12). We thank Scott Tremaine for drawing our attention 
to that work.}
Using~(\ref{eq:corrsp}),
this  gives 
\begin{equation}
\begin{split}
\langle (\Delta v_p)^2 \rangle =&  16 G^2 m 
\int d {\bf v} \frac{f ({\bf v})}{|{\bf v}_p \!-\! {\bf v}|}\\
& \int_0^T \! \frac{(T \!-\! t)}{t}  
\left[{\rm Si} (k_x |{\bf v}_p\! -\! {\bf v}| t) 
- {\rm Si} (k_m |{\bf v}_p\! -\! {\bf v}| t) \right] d t, 
\end{split}
\end{equation}
where we set ${\bf r} = {\bf r}_p= {\bf v_p } t$ to represent the time variation due to the test particle's motion 
(again, as in standard two body relaxation theory, this is rectilinear),
with ${\bf v}$  representing the motion of field particles. 
Proceeding as previously, while deriving equation (\ref{eq:tbs}), we find that in the 
diffusion limit --- which here requires that $k_m |{\bf v}_p - {\bf v}| t \gg 1$
is reached for all ${\bf v}$ --- the test particle's velocity dispersion increase, 
as it moves through the fluctuating field, can be expressed as  
\begin{equation}
\langle (\Delta v_p)^2 \rangle = 8 \pi m G^2 T \ln\left(\frac{k_x}{k_m}\right) \int d {\bf v} \frac{f ({\bf v})}{|{\bf v}_p - {\bf v}|}.  
\label{eq:velydist}
\end{equation}
This reduces to  (\ref{eq:tbs}) if $f ({\bf v}) = \rho_0 \delta_D ({\bf v}_f  - {\bf v})$, in which case the relaxation time 
is given by (\ref{eq:tbr}).

Finally, note that 
the velocity dispersion derived thus, in the diffusion limit,  is related to the trace of the diffusion coefficient matrix by 
$\langle (\Delta v_p)^2 \rangle 
= T \sum_{i = j} D [\Delta v_i \Delta v_j]$. 
The individual diffusion coefficients $D_{i j}$ can also be obtained in a similar manner
(as detailed in Appendix~\ref{app:diff}).

\section{Relaxation induced by fluctuating axion system} 
\label{sec:axion_fluc}

The same logic behind the derivations of the 
classical relaxation time, including  the 'sweeping' assumption used, can be employed 
to determine  the effect of a fluctuating FDM axion field on a classical test particle, with some  
differences arising from peculiarities  connected to the quantum origin 
of the evolution of the density fluctuations of the axion field. 

As FDM axions are by definition ultra light (with masses around $10^{-22} {\rm eV}$), 
an enormous number of them is needed to constitute a dark matter halo  ($\sim 10^{100}$). There is little question here of discreteness noise, 
the source of classical relaxation discussed above,  
having much effect; a mean field approximation is therefore warranted.
For a system of  bosons of mass $m$, interacting only through gravity,  this leads to the  
Schr\"odinger-Poisson system \citep{Ruffini1969}
\begin{equation}
i \hbar \frac{\partial}{\partial t} \psi ({\bf r}, t) = - \frac{\hbar^2}{2 m} \nabla^2 \psi ({\bf r}, t) + m~ \Phi_s ({\bf r}, t)  \psi ({\bf r}, t), 
\label{eq:schrod}
\end{equation}
\begin{equation}
\nabla^2 \Phi_s ({\bf r}, t) =  4 \pi G ~|\psi ({\bf r}, t)|^2,
\label{eq:PoiV}
\end{equation}
where $\Phi_s$ is the self consistent gravitational potential.

Since, for the system of bosons so described, there are many particles in the same state, 
$\psi$ behaves as a classical field. The square of the norm of the wave function is directly proportional 
to the number of particles around position vector ${\bf r}$ at time $t$, and if $\psi$ is mass normalized 
(as assumed above) then $|\psi|^2$ is the mass density.  This is stable when observed, 
in the sense that the result of observation is not subject to issues such as the collapse of the wave function. 
The quantum-origin of the dynamics, arising 
from the large de Broglie wavelength, manifests itself nevertheless through interference patterns and fluctuations
ubiquitous in numerical simulations of such systems \citep[e.g., ][]{Schive2014I}. The 
role of $\hbar$ is to set an effective spatial (and mass) scale for the fluctuations,  
given   axion masses $m$ and speeds $v$ (and mean density $\rho_0$). 
It is the effect of those 
fluctuations on a classical particle that is to be modelled using the methods developed above for 
discreteness noise. 

To mimic the derivation of the two body relaxation time, 
we again assume an infinite homogeneous medium, while neglecting self gravity of 
the axion field.  In other words we invoke the 'Jeans-Chandrasekhar swindle'
(see discussion in paragraph following equations (\ref{eq:phirk}) and (\ref{eq:rhork})). 
We do this 
by first setting $\Phi_s = 0$ in equation~(\ref{eq:schrod}). 
To describe the effect of fluctuations, in this context, we replace $\Phi_s$ in~(\ref{eq:PoiV})
with $\Phi ({\bf r}, t) = \Phi_s ({\bf r}, t) - \langle \Phi_s ({\bf r}, t) \rangle$.
In this case one can analyze the density and potential fluctuations as in equations 
(\ref{eq:phirk}) and (\ref{eq:rhork}),  and they will still be related by (\ref{eq:phikk}). 
As this neglects the self-gravity of the fluctuations, they must be much smaller 
than the Jeans length, which is of the order of the physical size in an actual inhomogeneous system. 
For the swindle to be valid therefore the de Broglie wavelength must be significantly 
smaller than the size of the system.  

Conservation still requires that the classical density modes $\rho_{\bf k}$  be 'swept' according to 
equation~(\ref{eq:sweeprho}).  This will be the case if 
\begin{equation}
\phi_{\bf k} (t) = \phi_{\bf k} (0)  e^{- i {\bf k}. {\bf v} t}.
\label{eq:qsweep}
\end{equation}
For, if
\begin{equation}
\rho ({\bf r}, t) = \psi ({\bf r}, t)  \psi^* ({\bf r}, t)     = \int \phi_{\bf k} (t) 
\phi^{*}_{{\bf k}^{'}} (t) e^{i ({\bf k} - {\bf k}^{'}) . {\bf r}} d {\bf k} d {\bf k}^{'} ,  
\end{equation}
and 
\begin{equation}
\rho_{\bf k}  (t) 
= \int  \psi \psi^* e^{i {\bf k} . {\bf r}} d {\bf r} = 
\int \phi_{{\bf k}^{'}} (t)  \phi^*_{{\bf k}^{'} - {\bf k}} (t) d {\bf k}^{'}. 
\end{equation}
Then 
\begin{equation}
\rho_{\bf k} (t) =  e^{- i {\bf k} . {\bf v} t}  \int \phi_{\bf k}^{'} (0) \phi^*_{{\bf k} - {\bf k}^{'}} (0) d {\bf k}^{'} = \rho_{\bf k} (0) 
e^{- i {\bf k}  . {\bf v} t} .
\end{equation}
On the other hand the free field Schr\"odinger equation
for the  axion system, 
\begin{equation}
i \hbar \frac{\partial}{\partial t} \psi = - \frac{\hbar^2}{2 m} \nabla^2 \psi,
\end{equation}
has a solution that can be Fourier expanded as    
\begin{equation}
\psi ({\bf r}, t) = \int  \phi_{\bf k} e^{i {\bf k}.{\bf r} - \omega t} d {\bf k}.  
\label{eq:SS}
\end{equation}
This can be rewritten as 
\begin{equation}
\psi ({\bf r}, t) = \int  \phi_{\bf k} (t) e^{i {\bf k}. {\bf r}} ~d {\bf k}, 
\end{equation}
with $\phi_{\bf k} (t)$ given by~(\ref{eq:qsweep}), provided 
${\bf k} . {\bf v}  = \omega$.  That is, if each mode is swept by
its  phase velocity.
As the solution~(\ref{eq:SS}) requires 
that 
\begin{equation}
\omega = \frac{\hbar k^2}{2 m}, 
\label{eq:dispersion}
\end{equation}
the group velocity $\frac{d \omega}{d {\bf k}} =  \frac{\hbar {\bf k}}{m}$
of a de Broglie wave packet is double the phase velocity.

In this way the dynamics 
of the free axion system and its effect on the motion of a classical test particle
can  be completely described, by invoking the 
sweeping assumptions of the previous sections and proceeding 
analogously. Note however an important difference.  
In the classical point particle case, a mode ${\bf k}$ 
could be swept by any velocity ${\bf v}$, since 
${\bf v}$ was independent of ${\bf k}$.  This is not the case here 
since
the velocities are directly dependent on 
the wave numbers through the  nonlinear dispersion 
relation~(\ref{eq:dispersion}).    

\subsection{Force correlation function and induced velocity variance}
\label{sec:forcecorr_ax}

In the context just set, 
the density contrast power spectrum can be written as 
\begin{equation}
\mathcal{P} ({\bf k}, t)\! = \!\frac{(2 \pi)^3}{\rho_0^2} \!\!\int\!\! d {\bf v_1} d {\bf v_2} f (\! {\bf v}_1\!) f (\!{\bf v}_2\!)
\delta_D ({\bf k} - m_\hbar\! {\bf v}_d) e\!^{-i  m_\hbar ({\bf v}_c . {\bf v}_d) t},
\label{eq:PS_Ax}
\end{equation}
where 
\begin{equation}
m_\hbar = 2 m/\hbar,
\label{eq:mhbar}
\end{equation}
${\bf v}_i = \hbar {\bf k}_i/ m$ are de Broglie wave packet group velocities, 
and ${\bf v}_c$ and ${\bf v}_d$ correspond to the sum and differences of the phase velocities
of interfering waves; such that 
$2 {\bf v}_c = {\bf v}_1 + {\bf v}_2$, 
and $2 {\bf v}_d = {\bf v}_1 - {\bf v}_2$, as detailed in Appendix~\ref{app:BOFT}. 
The above expression assumes 
that $\langle \phi_{\bf k} \rangle = 0$, $\langle \phi_{\bf k} \phi^{*}_{\bf k^{'}} \rangle = f_{\bf k} ({\bf k}) \delta_D ({\bf k} - {\bf k}^{'})$. 
The $\phi_{\bf k}$ are thus modes of a complex Gaussian random field. This is  
consistent with our assumption that the density field 
is  a homogeneous, stationary Gaussian random field, completely characterized by 
a power spectrum and two point correlation function,  
with the stochastic dynamics  determined by the force two point correlation function
(equation~\ref{eq:corrsp}; see also~\citealp{EZFC}).  Physically, this may also be justified 
if the waves are thought to arrive  at the test particle location from large distances and different directions 
with random phases (cf. \citealp{BOFT}), which is consistent with a random mode sweeping hypothesis. 
Furthermore,  we assume that the ${\bf k}$ space distribution function is related to the 
velocity  distribution of the axions by 
$f_{\bf k}  ({\bf k}_i) d {\bf k}_i =f ({\bf v}_{i}) d {\bf v}_{i}$,
and the distribution functions are mass normalised such that 
$\int f ({\bf v}_i) d {\bf v_i} = \int f_{\bf k} ({\bf k}_i) d {\bf k}_i = \rho_0$.

The equal time, spatial power spectra are thus
\begin{align}
\nonumber
\mathcal{P}({{\bf k}}, 0) & =
\frac{(2 \pi)^3}{\rho_0^2} \int d {\bf v_1} d {\bf v_2} f ({\bf v}_1) f ({\bf v}_2)
\delta_D ({\bf k} - m_\hbar {\bf v}_d)\\
& = \frac{(4 \pi)^3}{m_\hbar^3 \rho_0^2} \int d {\bf v}_c  f ({\bf v}_c + {\bf k}/m_\hbar) f ({\bf v}_c - {\bf k}/m_\hbar).
\label{eq:axion_onet}
\end{align}
These will be used below, along with the dispersion relation~(\ref{eq:dispersion}),
to roughly estimate the random force 
on a test particle due to fluctuations emanating from an FDM halo.

The  force correlation function resulting from Eq.~(\ref{eq:PS_Ax}) is given by
\begin{align} 
\nonumber
\langle {\bf F} (0, 0) .  {\bf F} (r, & t)\rangle = \\
& \left(\frac{4 \pi G}{m_\hbar}\right)^{2}
\!\!\! \int \! \frac{e^{i m_\hbar ({\bf v}_d . {\bf r}  - {\bf v}_c . {\bf v}_d t)}}{v_d^2}
f ({\bf v}_1) f ({\bf v}_2) d {\bf v}_1 d {\bf v}_2. 
\label{eq:fullqcorr}
\end{align}
Because this integral is highly oscillatory for large ${\bf v}_d$, one may suppose it is dominated by 
relatively small values of $v_d \ll  v_c$. If this is the case,  $f ({\bf v}_1)$ and $f ({\bf v}_2)$ can be approximated  
as $f ({\bf v}_c)$. 
This gives 
\begin{equation}
\frac{\langle {\bf F} (0, 0) . {\bf F} (r, t) \rangle}{ (8 \pi)^3 G^2 m_\hbar^{-2}}  = 
\!\int \!\! f^2 ({\bf v_c}) d {\bf v}_c \int
\frac{\sin \left(m_\hbar v_d |{\bf r}/t - {\bf v}_c| t\right)}{m_\hbar v_d |{\bf r}/t - {\bf v}_c| t} d v_d, 
\label{eq:axforccorr}
\end{equation}
since the Jacobian associated to the change of variables equals $8$. 

From equation~(\ref{eq:axion_onet}), it can be seen that this approximation is effectively 
equivalent to assuming the long wavelength limit of the spatial power spectrum, 
which then simply corresponds to white noise.  In this context, the integration over ${\bf v}_d$ 
is no longer weighed by convolution with the distribution functions, a situation which
can be partly corrected for by appropriate choice of the Coulomb logarithm appearing below.  This approach is 
particularly justified if the distribution function is roughly constant for smaller speeds (larger wavelengths) 
before sharply cutting off  beyond a characteristic value (filtering off smaller wavelengths), as in the case of a Maxwellian.  
In this case, we explicitly show (end of Section~\ref{sec:typical_dens}) that the mass fluctuations on large (relative to de Broglie) 
scales do tend to Poissonian noise, which puts in context the correspondence with the classical two body 
relaxation  that we now derive. 
   
As before, we then use equation (\ref{eq:corrsp}) and assume a test particle velocity ${\bf v}_p$
(recall that the test particle is classical)  to obtain 
\begin{align} 
\nonumber
\langle (\Delta v_p)^2 \rangle = &  ~2 (8 \pi)^3 G^2 m_\hbar^{-3}
\int d {\bf v}_c  \frac{f^2 ({\bf v}_c)}{|{\bf v}_p - {\bf v}_c|}
\int_0^T \frac{(T - t)}{t} \times \\
&\left[
{\rm Si} \left(m_\hbar v_{dx} |{\bf v}_p - {\bf v}_c| t \right)
- {\rm Si} \left( m_\hbar v_{dm} |{\bf v}_p - {\bf v}_c| t \right) \right] d t.
\end{align}
In the diffusion limit this gives, 
\begin{equation}
\langle (\Delta v_p)^2 \rangle = \left(\frac{4 \pi}{m_\hbar}\right)^3 8 \pi G^2 T   \ln \Lambda 
\int d {\bf v}_c  \frac{f^2 ({\bf v}_c)}{|{\bf v}_p - {\bf v}_c|},
\label{eq:ax_disp}
\end{equation}
where here
\begin{equation}
\Lambda = \frac{v_{dx}}{v_{dm}},
\label{eq:LFDM}
\end{equation}
is a ratio of maximal and minimal speeds, related (inversely) to associated wavelengths through the 
de Broglie relation. We evaluate and further discuss this Coulomb logarithm in specific cases 
in Appendix~\ref{app:Coul}. 

If we introduce
\begin{equation}
m_{\rm eff} = 
\left(\frac{4 \pi}{m_\hbar}\right)^3
\frac{\int f^2 ({\bf v}) d {\bf v}}{\int f ({\bf v}) d {\bf v}},  
\label{eq:meff}
\end{equation}
where $m_\hbar$ is given by (\ref{eq:mhbar}) and
\begin{equation}
f_{\rm eff} ({\bf v})  =  \frac{\int f ({\bf v}) d {\bf v}}{\int f^2 ({\bf v}) d {\bf v}}  f^2 ({\bf v}),  
\label{eq:feff}
\end{equation}
then equation (\ref{eq:ax_disp}) acquires the same form as~(\ref{eq:velydist}).

The effective mass and distribution functions  found above are as those 
in~\cite{BOFT}, where they enter into expressions for the diffusion coefficients.
Our derivation is different in its 
incorporation and extension  
of the idea of the sweeping of modes by a ${\bf k}$-dependent velocity field 
from turbulence theory, in order to obtain the space-time correlation function and power spectrum;
and in that it explicitly involves  
evaluation of the force fluctuations and associated correlation function, with the diffusion 
limit taken as a final step.  In the context of this formulation, it is clear that the 
effective mass and distribution function are associated  
with an assumption of white noise spatial power spectrum of density fluctuations, 
with minimal and maximal cutoffs
determined by a Coulomb logarithm.  
This explains the parallel with softened (but statistically homogeneous on large scales) classical systems, 
found by~\cite{BOFT}.  A central difference with classical particle systems, however, is encoded
in the  quantum-origin correlation of the sweeping velocity and wavenumber of the interfering 
waves, which sets the particular form of the effective quantities.

Again, as in the case of classical particles, in the diffusion limit, 
$\langle (\Delta v_p)^2 \rangle 
= T \sum_{i = j} D [\Delta v_i \Delta v_j]$. 
Also, the individual diffusion coefficients $D [\Delta v_i \Delta v_j]$ can be obtained as outlined in 
Appendix~\ref{app:diff}. 

Note however that, as opposed to the situation with equation~(\ref{eq:velydist}),
a delta function distribution in velocities
does not lead to an equation analogous to~(\ref{eq:tbr}), for the growth of 
velocity dispersion.  In particular the effective mass diverges, reflecting the fact that 
perfect knowledge of the velocity leads to absolute uncertainty in space.

\subsection{From density to force fluctuations and relaxation}

In this section we compare power spectra and correlation functions of  density fluctuations, 
calculated in the context of the model just presented, to published results from numerical simulations.  
We also present estimates of the typical force fluctuations connected to the
stochastic density field, thus characterized, and rough estimates of the associated
relaxation time of a classical test particle moving in the fluctuating force field.  

In order to compare with simulations we need to define a density and 
velocity distribution.  The assumption here  --- as in applications
of two body relaxation theory ---  is that our calculations are locally valid 
for inhomogeneous systems with local average density $\rho_0$. 

A realistic FDM halo is expected to follow the~\cite{nfw}  profile 
for radii significantly larger than the typical de Broglie 
wavelength of the FDM; that is beyond the   
solitonic core (\citealp{Schive2014}). 
In turn, the NFW profile can be well approximated in the intermediate radii 
$r$ around the maximal rotation speed (of order of the NFW scale length)
by an isothermal profile $\rho_0 \propto 1/r^2$ \citep{Chan2018}.  
Throughout this section we assume such a profile, along with a Maxwellian 
velocity distribution. Our calculations are therefore strictly valid only 
at radii larger than that of the solitonic core. 
We comment briefly on their possible relevance near and inside the core.

\subsubsection{Typical Density and mass fluctuations  in axion systems}
\label{sec:typical_dens} 

\begin{figure}
	\includegraphics[width=1 \columnwidth,trim={0.cm 0.1cm 1.2cm 1cm},clip]{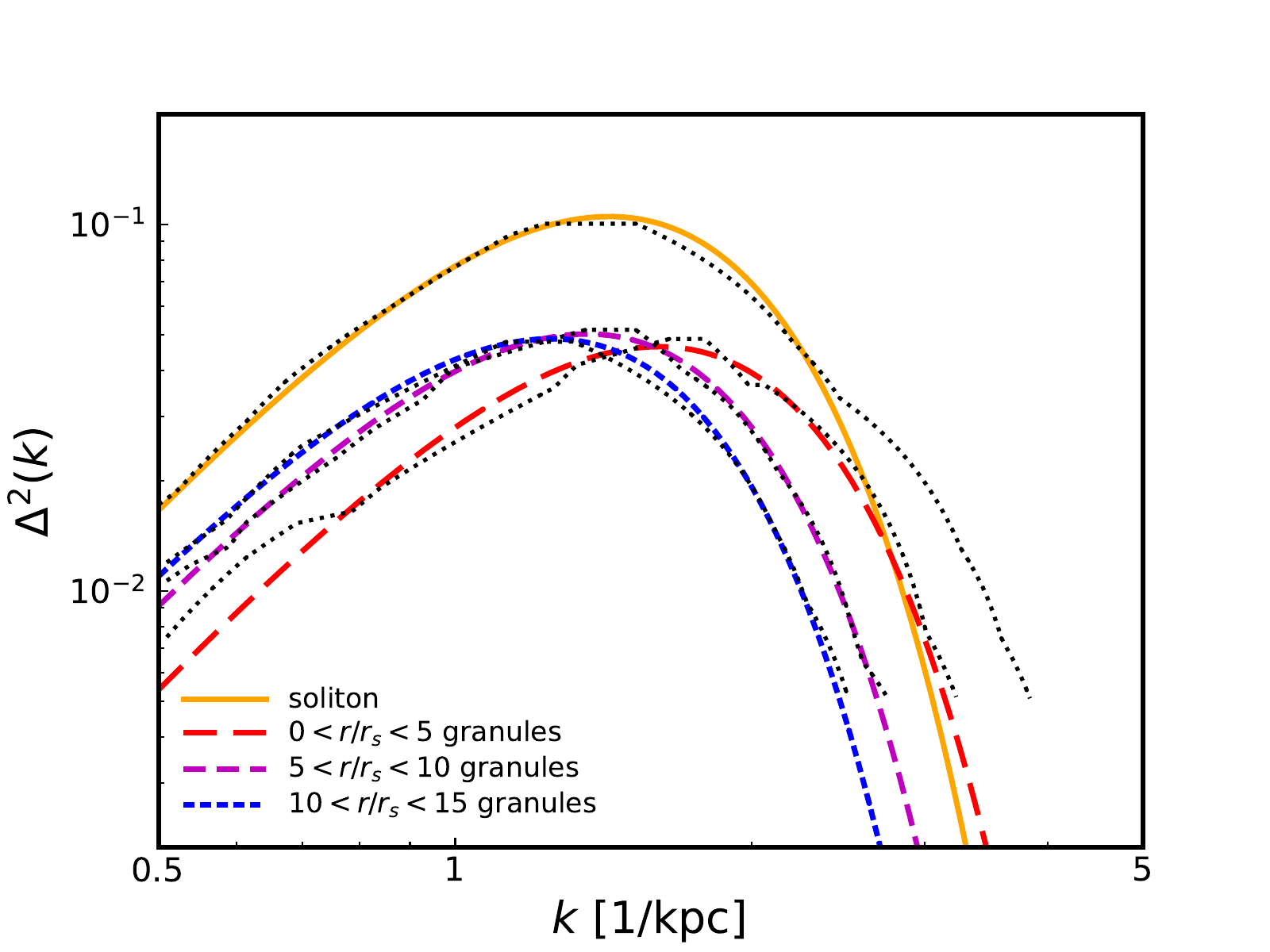}
	\caption{Equal time dimensionless power spectra of density fluctuations. The dotted lines represent 
(arbitrarily normalized) power spectra inferred from numerical simulation, 
as presented in Fig~4 (uppermost right hand panel) of~\protect \cite{Chan2018}. 
They correspond to power spectra taken within the soliton core $r_s$, and in bins $0 < r < 5 r_s$, $5 r_s < r < 10 r_s$  and $10 r_s < r < 15 r_s$.
The solid lines show best (least squares) fits using 
equations~(\ref{eq:eqtPSM}) and~(\ref{eq:DPS}). 
	}
	\label{fig:PS}
\end{figure}

\begin{figure}
     \includegraphics[width=1 \columnwidth,trim={0.cm 0.1cm 1.2cm 1cm},clip]{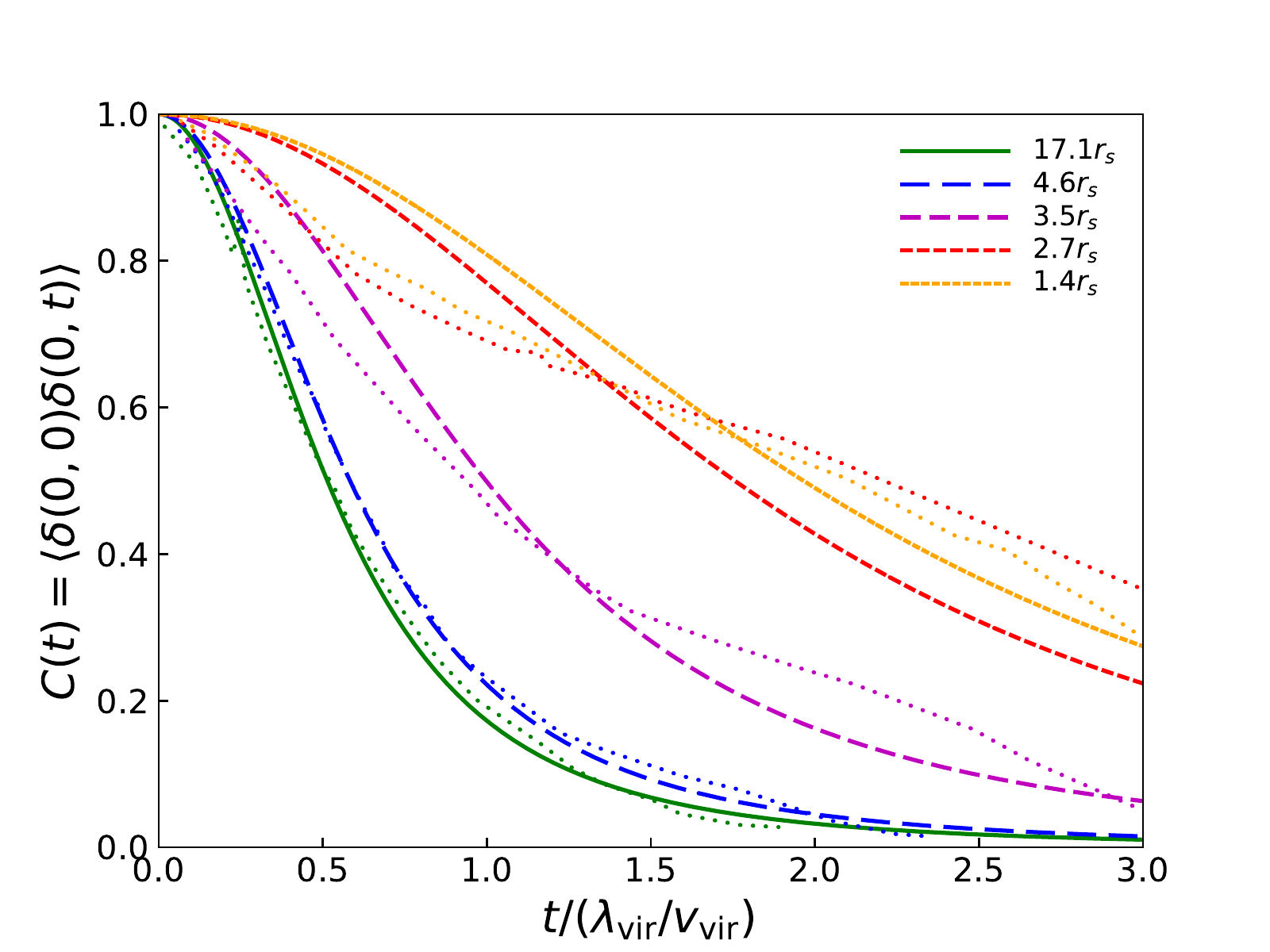}
	\includegraphics[width=1 \columnwidth,trim={0.cm 0.1cm 1.2cm 1cm},clip]{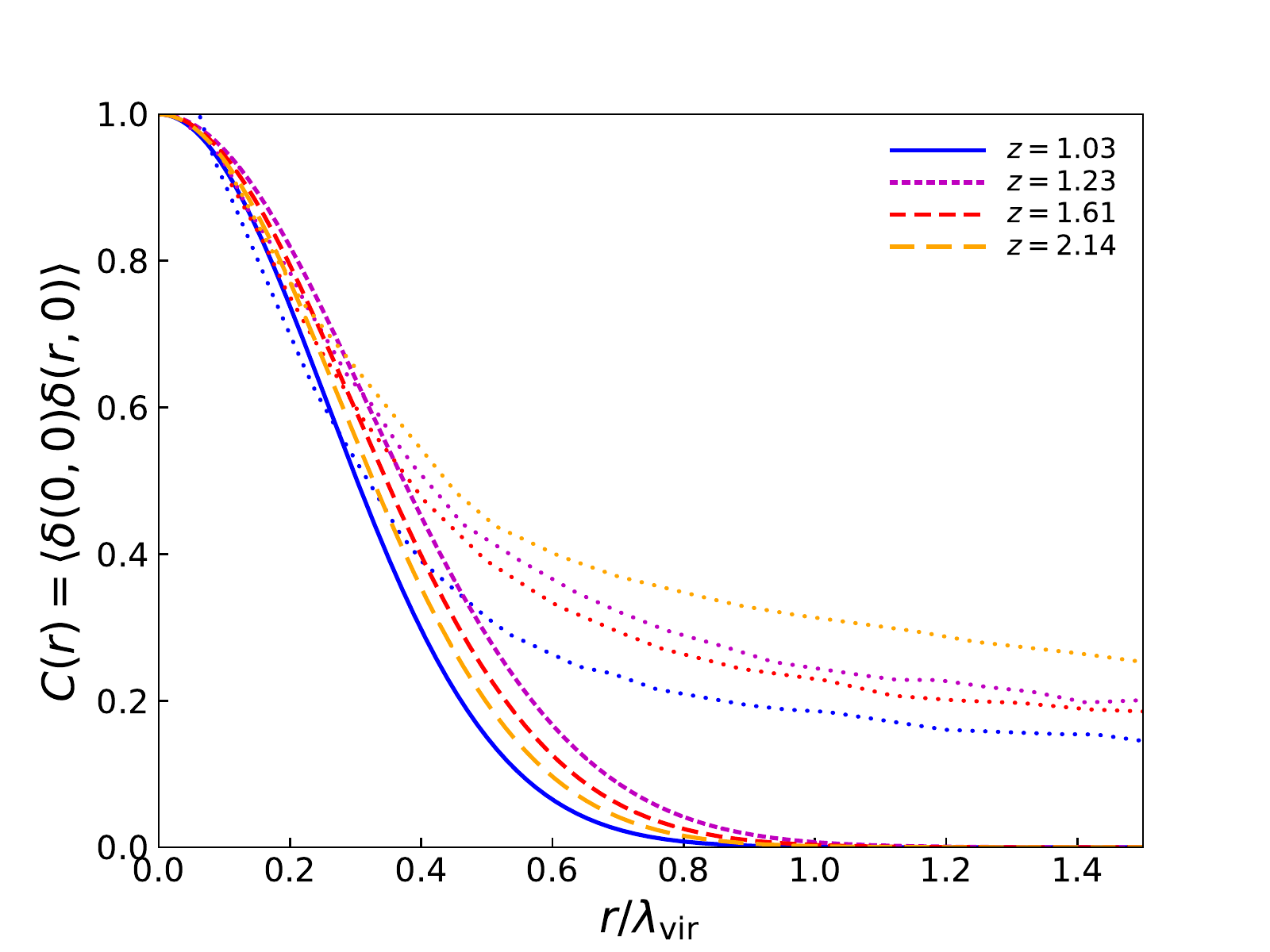}
	\caption{Correlation functions  of density fluctuations. The dotted lines show correlation functions  inferred 
from numerical simulations of~\protect \cite{Veltmaat2018} (their Fig.~8). 
The time correlation functions  (upper panel) are averaged over denoted multiples of soliton core radii $r_s$.
The solid lines are best (least squares) fits using equation~(\ref{eq:DencorrM}).
The lower panel shows the spatial correlation 
function at different redshifts. In the case of the spatial correlation functions, the 
fits take into account only points with $r/\lambda_{\rm vir} \le 0.4$. Here $\lambda_{\rm vir} = \frac{\hbar}{m v_{\rm vir}}$, 
where $v_{\rm vir}$ is virial velocity.  The effective best fit $\lambda$ to the spatial correlation 
function,  obtained by putting $t = 0$ in equation~(\ref{eq:DencorrM}), range 
from $\lambda = 0.36$ to $\lambda = 0.42$ in units of $\lambda_{\rm vir}$.   The fits are therefore good up to a scale
of order of the effective fitting wavelength. 
	}
	\label{fig:Corr}
\end{figure}

\begin{figure}
	\includegraphics[width=1 \columnwidth,trim={0.cm 0.1cm 1.2cm 1cm},clip]{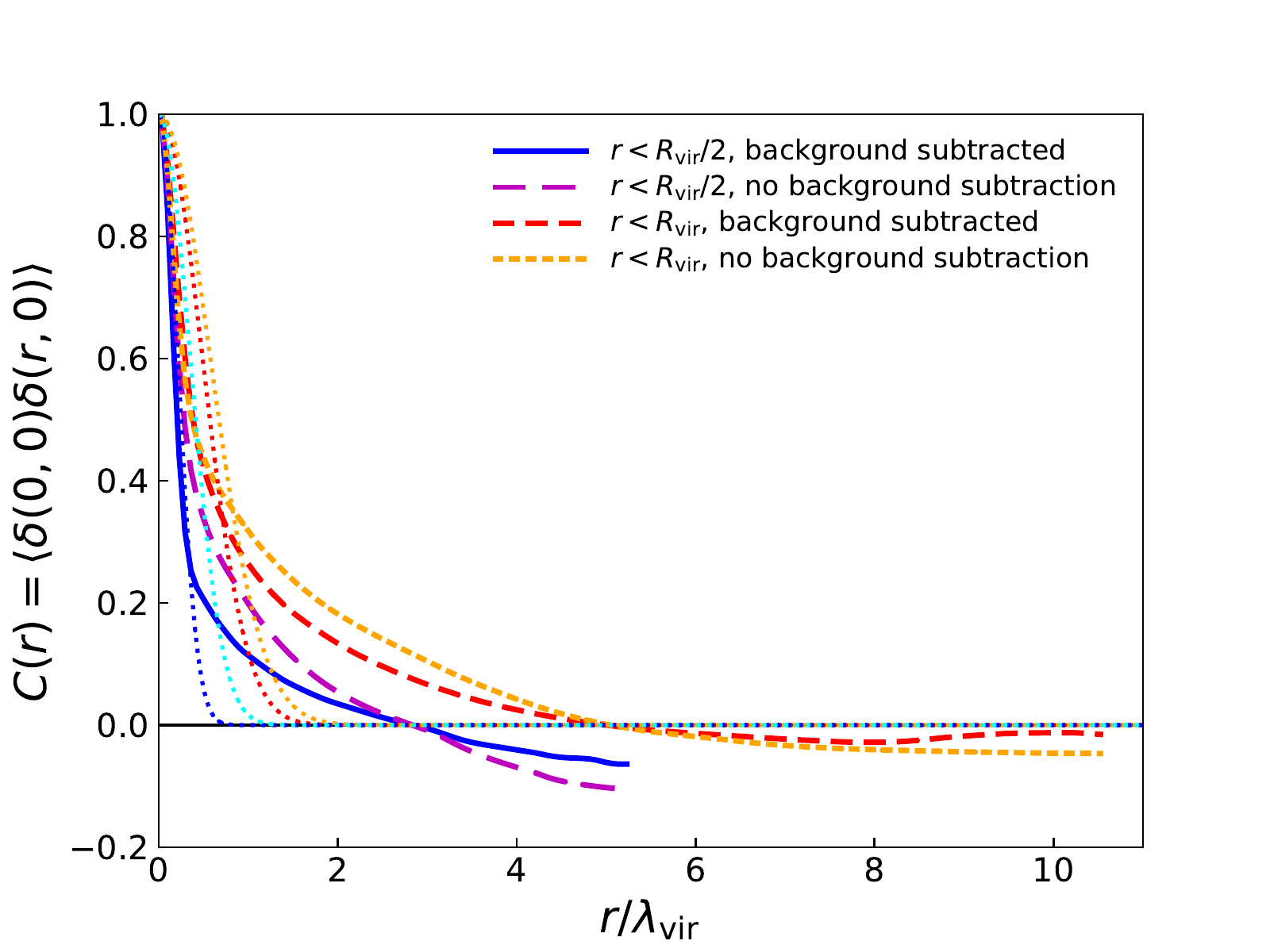}
	\caption{Spatial correlation functions of simulated haloes, evaluated inside the virial radius and half 
that radius, with and without subtracting the radially averaged halo background density  (simulation results kindly made available by Jan Veltmaat). 
The corresponding best fits using equation~(\ref{eq:DencorrM}) are also shown (dotted lines). 
These are good up to the scales of order of the effective fluctuation scales, and are better 
inside half the virial radius and when the background density is subtracted.  The long correlation tail may reflect the finite spatial size of 
the fluctuations, non-sphericity of the background halo density, or a radially varying de Broglie wavelength in a realistic halo. 
The effective wavelengths of the  fits (in units of of $\lambda_{\rm vir} = \frac{\hbar}{m v_{\rm vir}}$)  
are $\lambda_\sigma =  0.69, 0.81$ for correlation functions evaluated inside the virial 
radius, with and without background subtraction respectively, and $\lambda_\sigma = 0.3, 0.5$ 
inside half the virial radius, with and without subtraction respectively.
	}
	\label{fig: long}
\end{figure}

Since, in the case of axion systems, the wave number and velocity distributions
are necessarily related, the fluctuation power spectrum  
is determined by the velocity distribution. If we assume a Maxwellian distribution
with one dimensional dispersion $\sigma$,
\begin{equation}
f (v) = \frac{\rho_0}{(2 \pi \sigma^2)^{3/2}} e^{-\frac{v^2}{2 \sigma^2}}, 
\label{eq:Max}
\end{equation}
the equal time power spectrum given by 
equation~(\ref{eq:axion_onet}) becomes (cf. Appendix~\ref{app:max})
\begin{equation}
\mathcal{P}({{\bf k}}, 0) = \left(\frac{2 \sqrt{\pi}}{{m_\hbar} \sigma}\right)^3 e^{-\frac{k^2}{\sigma^2 {m_\hbar^2}}}. 
\label{eq:eqtPSM}
\end{equation}
The dimensionless power spectrum 
\begin{equation}
\Delta^2 (k) = \frac{k^3}{2 \pi^2} \mathcal{P}(k),
\label{eq:DPS}
\end{equation}
measures the contribution to the variance in density fluctuations from logarithmic bins around wave number $k$.

In Fig.~\ref{fig:PS}, we compare what is obtained from this formula with the arbitrarily normalized power spectra from numerical simulations of\cite{Chan2018}, who isolated dwarf sized FDM axion haloes 
from cosmological simulations solving the Schr\"odinger-Poisson system (\ref{eq:schrod} and \ref{eq:PoiV})  self consistently. 
(The haloes were then evolved along with classical particles representing a stellar distribution.
The relevant simulation results for us are those before this latter component is introduced.)
As can be seen, the fits are quite good, with some exceptions.  Notably at high wavenumbers, inside the solitonic core, where our 
Maxwellian cutoff at high wavenumber/velocities, appropriate for an isothermal sphere, may not be valid. 
Outside the core, the peaks move only slightly and the fits thus indeed correspond to a nearly 
isothermal system. Our {\it a priori} assumption of an isothermal system is thus approximately valid 
outside the core.  

Again assuming the Maxwellian distribution of  equation~(\ref{eq:Max}), and defining an  
associated wavelength $\lambda_\sigma = \hbar/ m \sigma = 2/m_\hbar \sigma$ 
the correlation function of the density contrast is found 
(by Fourier transforming~\ref{eq:PS_Ax}, cf. Appendix~\ref{app:max}) 
to be
\begin{equation}
\langle \delta (0, 0) \delta(r, t) \rangle = \frac{1}{(1+ \sigma^2 t^2/\lambda_\sigma^2)^{3/2}} e^{- \frac{r^2/\lambda_\sigma^2}{1 + \sigma^2 t^2/\lambda_\sigma^2}}.
\label{eq:DencorrM}
\end{equation}
Note that $\langle \delta^2 (0, 0) \rangle = 1$; the density fluctuations on the smallest 
scales are therefore of order unity.
The variance over all $k$ in density fluctuation contrast, given by
\begin{equation}
\langle \delta^2 \rangle = \frac{1}{(2 \pi)^3} \int_{k_m}^{k_x} \mathcal{P} (k) 4 \pi k^2 d k = \int_{k_m}^{k_x} \Delta^2(k) d\ln k, 
\end{equation}
 also tends to unity as $k_m \rightarrow 0$ and $k_x \rightarrow \infty$.

Fits, using  equation~(\ref{eq:DencorrM}), 
to correlation functions presented by~\cite{Veltmaat2018} (their figure 8) 
are shown in Fig.~\ref{fig:Corr}.   
The numerical results of~\cite{Veltmaat2018}
are based on  cosmological 
simulations using a standard $N$-body code  
for most of the simulation volume, with high resolution zoom in at selected halo locations solving the 
Schr\"odinger-Poisson system.  Boundary conditions at the `Schr\"odinger domain' 
are imposed {\it via} a wave function evolved according to the Hamilton-Jacobi equation.

The fits to the time correlation functions inferred from simulations are good,
especially given that they depend on a single parameter. 
 The fits are again better at larger radii. This may be expected, as in addition to 
our assumption of Maxwellian velocities,  the assumption that 
the fluctuation modes  are randomly 'swept' and come with random phases may not be quite valid at smaller radii.
Indeed, as can be seen, at smaller radii the time-correlations display long tails, possibly signaling that the 
fluctuations are progressively  harder to describe by the sweeping of a  fluctuating field  that corresponds to 
white noise beyond the de Broglie scale.  Though~\cite{Veltmaat2018} find persistent oscillations of order $30 \%$ 
even inside the core, these come with a characteristic frequency, with limited spread around it (their figure 7), 
If not entirely deterministic, the quasi-coherent core oscillations  may 
still have an effect on the stellar dynamics of central bulges and star clusters of the sort described here. 
This may be further  enhanced by coupling to  baryonic energy feedback leading to stochastic density and potential fluctuations, 
(as the oscillations correspond  to long lived excitations from the solitonic ground 
state).  But the description of their effect would require modification of the present formulation, in order to take into account
the limited frequency spread (we further comment on this in  Appendix~\ref{App:Eri}).

More puzzling is the behaviour of the spatial correlation function,
which displays a quite weak decay tail at large separations.  
Jans Veltmaat was kind enough to provide correlation functions  up to larger scales, 
with and without radial background density subtracted (shown in Fig.~\ref{fig: long}). 
The results are better fit by our model in the former case. Nevertheless, significant 
discrepancy remains at $r \ga 0.5 \lambda_{\rm vir}$, that is at 1 to 1.5 effective 
wavelengths defined by the best fitting $\lambda_\sigma$ (from equation~\ref{eq:DencorrM} with $t = 0$). 
The correlation tail may be explained if background 
density correlations remain after the subtraction of the radial averaged density; e.g.,  because the haloes are 
not in fact spherical. Indeed, the long tail of the spatial correlation functions tend to  mimic those of a purely 
cold dark matter halo simulated in tandem (Jan Veltmaat, private communication).  
In addition, the effective size of the the FDM is not small compared to the 
radial binning, and this  'size' is not constant but varies with radius, due to changing velocity (and thus de Broglie wavelength) in a realistic halo. 
The effective size reflected in the $\lambda_\sigma$ of our fits thus necessarily reflects an averaging. The fact that 
our fits are better inside half the virial radius may point to effects of changing velocity dispersions
not taken into account by our isothermal model.  
 
The long correlation tail may appear to contradict the much better fits we obtained to the 
fluctuation power spectra of Chan et al. 
These,  as we saw above, are consistent with our  model. 
At larger scales they generally correspond to a white noise power spectrum 
(flat $\mathcal{P} (k)$ with $\Delta^2 \sim k^3$).  This agreement could
be because the power spectra do not go to large enough
scales for discrepancies due to the background density correlations to appear. 
Also, except for the red line in Fig~1, which indeed shows some excess power on larger scales, 
the Chan et al. spectra are  evaluated over radial shells rather than inside spherical bins, which leaves less room 
for effects due to background density and velocity profiles (although the same general form of the power spectra 
also arises for the self consistent equilibrium haloes built by~\citealp{Lin_Chiueh18}, where
spherical bins were used; their figure 7c).

If not an artifact arising from the definition and subtraction of the background density, 
the spatial correlation function 
of Veltmaat et al. would imply large scale correlations  that cannot be captured by the calculated correlation functions --- 
at least while keeping to isothermal Maxwellian systems ---   
as the model  predicts mass fluctuations that decay as Poisson noise beyond the 
de Broglie wavelength, consistent with the white noise power spectrum at larger scales.  
This can be explicitly shown by calculating 
the variance a filtering scale $R$, which is given by 
\citep[e.g., ][]{Martinez2002, Mo2010}
\begin{equation}
\label{eq:var}
\sigma_R^2 = \frac{1}{2 \pi^2}  \int_0^\infty W^2(k,R) \mathcal{P}(k) k^2 dk,
\end{equation}
where $W$ is the  Fourier transform of the window filtering function.  
Using a Gaussian filter, such that  $W^2(k,R) = e^{-R^2 k^2/2}$ gives,
\begin{equation}
\sigma_R^2 = \left( \frac{2}{(R \sigma m_\hbar)^2 + 2} \right)^{3/2},
\end{equation}
which shows the mass variance to be of order one on scales smaller than the typical de Broglie wavelength of
FDM particles and decreasing as $1/R^3$ on larger scales, as expected of  Poisson noise.

\subsubsection{Force fluctuations and rough estimates of the relaxation time}
\label{sec:force_fluc}

The mean square of the force fluctuations can be obtained from equation~(\ref{eq:axforccorr})
\begin{equation}
\langle F^2 (0, 0) \rangle =( 8 \pi)^3 G^2 m_\hbar^{-2} v_{d x} \int f^2 ({\bf v}) d {\bf v},
\end{equation}
where we have assumed that $v_{d x} \gg v_{d m}$. 
Recalling that $\int f ({\rm v}) d {\bf v} = \rho_0$, Eq. ~(\ref{eq:meff}) yields
\begin{equation}
\langle F^2 (0, 0) \rangle = 8~G^2 \rho_0 m_\hbar v_{d x} m_{\rm eff}.
\end{equation}
If $v_{ d x} \approx \frac{\hbar k_{{x}}}{2 m}$, and  the maximum wave number is related to the minimal 
wavelength by $k_{{x}} = 2 \pi / \lambda_{{x}}$, then 
\begin{equation}
\langle F^2 (0, 0) \rangle = 16~\pi G^2 \rho_0 \frac{m_{\rm eff}}{\lambda_{\protect x}}.
\label{eq:for_fluc}
\end{equation}

A quite crude, but instructive,  estimate of the growth of the mean squared speed and associated relaxation time can then be 
obtained from assuming that the diffusion process determining the growth can be described by a collection of 
independent 'kicks', each of duration $\Delta t$, such that after $n$ kicks
\begin{equation}
\langle (\Delta v_p)^2 \rangle  \approx n \langle F^2 (0, 0) \rangle (\Delta t)^2.
\end{equation}
Or, using equation~(\ref{eq:for_fluc}), 
\begin{equation}
\langle (\Delta v_p)^2 \rangle  \approx \frac{8\pi G^2 \rho_0 m_{\rm eff}}{v_r} T.
\label{eq:disp_qc}
\end{equation}
Here $T = n \Delta t$ and we assumed a typical test particle velocity relative to the axions system modes  $v_r$ and that 
$\Delta t$ is equal to a half-mode modulation time of the minimal wavelength, such that
$\Delta t = \frac{\lambda_{\rm min}}{2 v_r}$. 
Although equation (\ref{eq:disp_qc}) is the same as (\ref{eq:tbs}) --- if   
$\ln \Lambda = 1$ and $m$ replaced by the effective mass ---  
as noted above, a velocity distribution of FDM axion velocities must 
be defined if  this $m_{\rm eff}$, and consequently  
$\langle (\Delta v_p)^2 \rangle$, does not diverge.  
For the Maxwellian distribution (\ref{eq:Max}) 
\begin{equation}
m_{\rm eff} = \frac{8 \pi^{3/2} \rho_0}{m_\hbar^3  \sigma^3}. 
\label{eq:meffM}
\end{equation}

A better estimate of the increase in velocity 
dispersion due FDM density and force fluctuations 
can be obtained by calculating the diffusion coefficients
calculated in Appendix~\ref{app:diff}.
In this way
\begin{equation}
\langle (\Delta v_p)^2 \rangle  = 
T \sum_{i = j} D [\Delta v_i \Delta v_j]
= T \left( D [(\Delta v_\parallel)^2] + D [(\Delta v_\perp)^2] \right).
\label{eq:todiff}
\end{equation}
For a Maxwellian velocity distribution this gives
\begin{equation}
\langle (\Delta v_p)^2 \rangle
=  T \frac{\sqrt{2}~4 \pi G^2 \rho_0 m_{\rm eff} \ln \Lambda}{\sigma_{\rm eff}} \left[\frac{{\rm erf}  (X_{\rm eff})}{X_{\rm eff}}  \right].
\label{eq:max_disp}
\end{equation}
Here $\sigma_{\rm eff} = \sigma/\sqrt{2}$ and $m_{\rm eff}$ is given by 
(\ref{eq:meffM}) and $X_{\rm eff} = v_p/(\sqrt{2}\sigma_{\rm eff}) = v_p/\sigma$.
Thus we have 
\begin{equation}
\langle (\Delta v_p)^2 \rangle 
= T
\frac{8 \pi G^2 \rho_0 m_{\rm eff} \ln \Lambda}{v_p}  {\rm erf}  (X_{\rm eff}).
\label{eq:veldispt}
\end{equation}
If $v_p \sim v_r$ and is set to  $\sqrt{3} \sigma$ then ${\rm erf}  (X_{\rm eff}) = 0.99$, and 
with replacement of $m$ with the appropriate $m_{\rm eff}$, 
this expression is again virtually equivalent to (\ref{eq:tbs}). The time for the fluctuations to induce
a velocity dispersion of the order of $v_p$ is accordingly also given by~(\ref{eq:tbr}), with $m_{\rm eff}$ 
in the denominator.

\section{The heating of disks by FDM axions}
\label{sec:disk}

\subsection{First estimate}

If  one wishes to obtain a rough estimate of the 
role  of the fluctuating force in FDM haloes in significantly increasing
the velocity dispersion of embedded disk stars,  then $v_p$ in 
equation~(\ref{eq:veldispt}) can be replaced by the 
circular velocity $v_{\rm circ}$ and again  ${\rm erf}  (X_{\rm eff})  \approx 1$. 
Without taking into account the disk's self gravity, the relaxation timescale, 
taken to produce a velocity variance in the motion of the disk stars $\sigma_d^2$ 
($\sigma_d$ being the dispersion), is
\begin{equation}
t_r \approx \frac{\sigma_d^2 v_{\rm circ}}{8 \pi G^2 \rho_0 m_{\rm eff}  \ln \Lambda},
\label{eq:tbrd}
\end{equation}
where $\sigma_d^2 = \sigma_z^2 +  \sigma_R^2 + \sigma_\theta^2$, when measured in cylindrical coordinates 
moving at the local circular velocity.   
To estimate this  timescale,  we assume the FDM halo outside the core to be represented by an isothermal 
distribution with radial mass density
\begin{equation}
\rho_0 = \frac{\sigma^2}{2 \pi G r^2},
\label{eq:diso}
\end{equation}
with associated circular speed $v_{\rm circ} = \sqrt{2} \sigma$. For this  density distribution
\begin{equation}
m_{\rm eff} = \frac{4\sqrt{2 \pi}}{G m_\hbar^3 v_{\rm circ} r^2}.
\label{eq:meffMiso}
\end{equation}   
Using  equations (\ref{eq:tbrd}), (\ref{eq:diso}) and (\ref{eq:meffMiso})  we thus find for 
for the Milky Way solar neighbourhood
\begin{equation}
t_r = \frac{1.1 \times 10^{12}} {\ln \Lambda} \left(\frac{r}{8~{\rm kpc}} \right)^4   \left(\frac{m}{10^{-22} {\rm eV}} \right)^3   
\left(\frac{\sigma_d}{70~{\rm km s^{-1}}} \right)^2 {\rm yr}.
\label{eq:tesimple}
\end{equation}
In Appendix~\ref{app:Coul} we estimate the Coulomb logarithm argument as 
\begin{equation}
\Lambda  = 36.9 \left(\frac{r}{8 {\rm kpc}} \frac{m}{10^{-22} {\rm eV}} \frac{v_{\rm circ}}{200 {\rm km/s}}\right)^{1/2}, 
\end{equation}
so that $\ln \Lambda = 3.6$. 

Thus for solar neighbourhood parameters this simple estimate suggests that the 
mass of the FDM axion should not be less than $\sim 0.36 \times 10^{-22} {\rm eV}$ 
if the local velocity dispersion resulting from fluctuation arising from FDM heating
through a Hubble time is not to exceed that observed.  

 It is remarkable that the
estimate of the relaxation time in equation~(\ref{eq:tesimple}) does not depend on 
the disk circular velocity.  On the other hand, observations suggest a clear 
 correlation between velocity dispersion in disks and their maximal rotation speeds
\citep{Bott93, KVdK, KVdKFr}. We also note that equation~(\ref{eq:tesimple})   
implies a velocity dispersion increase that 
scales in time diffusely  as $\sim t^{1/2}$. As we will see below this applies to both vertical and radial 
velocity dispersion.   
However the latter is observed to  scale as $\sim t^{1/3}$. 
Such discrepancies imply that only a small fraction of the velocity dispersion in disks 
can result from FDM axion fluctuations as treated here. Otherwise, the aforementioned 
scalings will not be reproduced.

\subsection{Vertical and radial dispersion and disk response} 
\label{sec:verad}

The estimate just presented, of the increase in disk star velocity dispersion, did not distinguish between 
vertical and radial increase. It also did not take into account, even in an approximate manner, the 
disk self gravity. To make progress on such issues we make use of the 
formulation of Binney \& Tremaine (2008, Section 7.4), exploiting the fact 
that the effect of fluctuations in FDM axion haloes is expected to be effectively  
equivalent to that of quasi-particles of mass $m_{\rm eff}$ and distribution function $f_{\rm eff}$. 

In this context, the rate  of  energy transfer, per unit mass, 
to disk stars in the vertical direction is 
\begin{equation}
D [\Delta E_z] =  D \left[\Delta \left(\frac{1}{2} v_z^2\right) \right] = v_z D [\Delta v_z] + \frac{1}{2} D [(\Delta v_z)^2].
\end{equation}
Assuming again a Maxwellian FDM velocity distribution and stellar motion with 
circular velocity $v_{\rm circ}$,  writing the diffusion coefficients in terms of components parallel and 
normal to the motion (Appendix~\ref{app:diff}), and ignoring terms suppressed by factors $(v_z/v_{\rm circ})^2$, 
one finds
\begin{align}
\nonumber
D [\Delta E_z] & = \frac{1}{4} D [(\Delta v_\perp)^2]\\
& = \frac{\sqrt{2} \pi G^2 \rho_0 m_{\rm eff} \ln \Lambda}{\sigma_{\rm eff}} \left[\frac{{\rm erf}  (X_{\rm eff}) - {\rm G_{\rm eff}} (X_{\rm eff})}{X_{\rm eff}}  \right],
\end{align}
where $m_{\rm eff}$ and $f_{\rm eff} (v)$  are given by (\ref{eq:meff}) and (\ref{eq:feff}) and
$X_{\rm eff} = v_p/(\sqrt{2}\sigma_{\rm eff}) \approx v_{\rm circ}/\sigma$, for nearly circular 
orbits. Then, using equation (\ref{eq:diso}) and  $v_{\rm circ} = \sqrt{2} \sigma$, 
\begin{equation}
D [\Delta E_z] = 0.39 \frac{G m_{\rm eff} v_{\rm circ}}{r^2} \ln \Lambda. 
\label{eq:DEz}
\end{equation}
If one assumes the virial relation for a system of self gravitating sheets to approximate 
a disk system $E_z$ is related to the vertical velocity dispersion 
$E_z = \frac{3}{2} \sigma_z^2$ (alternatively, the epicyclic approximation  leads to $E_z = \sigma_z^2$; \cite{LO1985}). 
Thus $d \sigma_z^2/d t = \frac{2}{3} D [\Delta E_z]$.  Integrating this, and using (\ref{eq:DEz}) and (\ref{eq:meffMiso}) 
one finds
\begin{equation}
\sigma_z = 2.4~{\rm km/s}\\ \left(\frac{10^{-22}{\rm eV}}{m} \right)^{3/2} \left(\frac{8 {\rm kpc}}{r} \right)^2 
\left(\frac{T}{10 {\rm Gyr}}\right)^{1/2} \ln \Lambda^{1/2} 
\label{eq:sigZ}
\end{equation}
Note again that the results do not depend on the circular speed associated with the isothermal halo.

As in the previous subsection, if we consider the vertical dispersion of stars to arise solely from
the FDM axion fluctuations  then the above leads to the relatively weak constraint of $m \ga 0.3 \times 10^{-22} {\rm eV}$, 
if the maximal vertical velocity dispersion in the solar neighborhood taken to be about $30 {\rm km/s}$. 

This constraint can be tightened by considering the radial dispersion. A similar analysis to that 
outlined above (again. following the calculation of the aforementioned section of Binney \& Tremaine)
shows that $\sigma_R = 1/0.53 \times \sigma_z$. Therefore 
\begin{equation}
\sigma_R =  4.5~{\rm km/s} \left(\frac{10^{-22}{\rm eV}}{m} \right)^{3/2} \!\left(\frac{8{\rm kpc}}{r} \right)^2 
\!\left(\frac{T}{10 {\rm Gyr}}\right)^{1/2} \!\!\ln \Lambda^{1/2} 
\label{eq:sigR}
\end{equation}
This is inconsistent with observations, as it predicts
$\sigma_R \sim \sigma_z \sim t^{1/2}$,  
while a power law more akin to $\sigma_R \sim t^{1/3}$ is suggested 
by the observations, a result exhaustively confirmed  by the recent data of~\cite{Mack_Bovy2019}. 

The origin of the observed age velocity dispersion relation in discs is not entirely understood. 
Contributions traditionally considered in the literature include  secular processes arising from scattering 
of stellar trajectories by molecular clouds 
and spiral arms (cf. Section~5,2 of~\citealp{Mack_Bovy2019} for a discussion of their possible relative contribution 
in light of their data).  The correlation could, on the other hand, reflect the formation history of a once gas rich
disk, rather than secular evolution (e.g.~\citealp{Bournaud2009, Vincenzo2019}).
If we assume that, due to the lack of proper correlation in $\sigma_R (t)$, 
 FDM fluctuations are marginal in deciding
the disc velocity dispersions,    and that therefore 
only the errors in their $\sigma_R$ can be tolerated as being due 
to FDM fluctuations, this is equivalent to setting $\sigma_R \sim 3 {\rm km/s}$ in~(\ref{eq:sigR}), which corresponds to 
$m \ga 2 \times 10^{-22} {\rm eV}$.

\subsection{Comparison with other work}
\label{sec:comparison}

\cite{Hui_etal2017} estimate the effect of FDM fluctuations on disk thickness by assuming
that they can be modelled as due to classical particles with effective mass  corresponding to that 
enclosed within half a de Broglie wavelength $\lambda_{\rm DB}$. Their equation  (35)  implies that the
effect is entirely dominated by encounters with minimal impact parameters 
$b_{\rm min}$, 
without the usual two dimensional 
integration over impact parameters that leads to the Coulomb logarithm. The notion that adiabatic fluctuations 
do not contribute to stochastic increase in velocity dispersion envisioned in standard relaxation theory
(cf. Appendix~\ref{app:Coul} and \citealp{Church_2018}), is taken into account by dividing this minimal 
$b_{\rm min}$ by the disk half-thickness.  
Their  equation (37), although strictly valid for axion masses $> 10^{-22} {\rm eV}$, gives constraints on the FDM mass 
from disk velocity dispersion in the solar 
neighbourhood  that are only slightly weaker than those we find here, where we have evaluated the combined 
dynamical effect  of the full spectrum of contributing Fourier modes (which leads to the Coulomb logarithm 
evaluated in Appendix~\ref{app:Coul}). 
Application of their equation~(36) on the other hand, relaxes the constraints on the mass much more significantly.

The effect of FDM fluctuations on the vertical dispersion of galactic disks were 
 also discussed by~\cite{Church_2018}. When we assume, as they do, that all 
the vertical disk dispersion is due to FDM halo fluctuations, we get  a similar but weaker constraint
of $\sim 0.3 \times 10^{-22} {\rm eV}$ on the FDM particle mass (instead of their $0.6 $).  
There are several possible reasons that can account for this difference.  
As that work was largely concerned  with heating due to classical subhaloes, the FDM fluctuations are derived by 
assuming classical particles of effective mass $M_\omega$ (their Eq. 18), and 
directly applying standard two body relaxation theory. 
The diffusive effects are also not resolved into vertical and parallel components
(associated with the vertical and parallel diffusion coefficients derived here).   
Furthermore, the effective mass of the FDM quasiparticles is $6.7$ times our $m_{\rm eff}$
(note that there seems to be a typo in their Eq. 24a, with a factor of $\sqrt{2}/4$ apparently missing). Under their assumptions, 
the Coulomb logarithm is also expected to be larger than evaluated here (Appendix~\ref{app:Coul}). 
Finally, their formulation of the disk response (their equation 25) is different in two ways than 
assumed here:  the factor 
$\kappa$  in the second term is in our case $1/3$ instead of 0.52 (cf. discussion following 
Eq.~\ref{eq:DEz} above), and their formulation includes an extra term 
describing the effect of mass accretion, which we ignore. 
This term is always positive, and thus increases further the effect of FDM fluctuations, but the other difference  
mentioned above can account for the discrepancy of a factor of about 
two in FDM mass constraint. 

When, by noting  that the increase in disk dispersion due to FDM fluctuations violates the scaling
relations between $\sigma_R$ and  $v_{\rm circ}$, and its time dependence 
(predicted as $\sigma_R  \sim t^{1/2}$, instead of the observed $\sim t^{1/3}$), we can derive tighter constraints 
on the mass of the FDM axion. Namely $m \ga 2 \times 10^{-22} {\rm eV}$.  This is similar  
to that derived by \cite{AmLoeb2018} from the effect of fluctuations on the dynamics of stellar streams in the Milky Way.  It is significantly 
weaker however than that found by~\cite{Marsh2018} by applying the fluctuation model of~\cite{EZFC} 
to the central star cluster ultrafaint Dwarf Galaxy Eridanus II.

As shown in Appendix~\ref{App:Eri}, directly applying our extended and improved model 
to the dynamics central cluster of Eridanus II leads to the similarly strong constraint
$m \ge 8.8 \times 10^{-20} {\rm eV}$, given one assumes that the FDM makes up all the 
dark matter and the effect associated fluctuations goes entirely 
into expanding the cluster, in the manner envisaged by~\cite{Marsh2018}. 
However, for masses $m \ga 10^{-20} {\rm eV}$ the cluster 
should lie inside the solitonic core of the FDM distribution~(\citealp{Marsh2018}). 
Strictly speaking, one can thus not rule out masses below this using the methods 
presented here; the eliminated FDM mass range is thus limited
(to $10^{-20} {\rm eV} \la  m   \la 10^{-19} {\rm eV}$), and does not include 
the range most interesting for solving  galactic scale problems of CDM. 

To extend the constraint on $m$ to lower values,
\cite{Marsh2018}  consider diffusion due to FDM central soliton core oscilations.
It may also be the case  that fluctuations from FDM granules outside the core could still 
affect the evolution of a cluster stars inside it.  However, as discussed in the aforementioned appendix, 
even leaving aside doubts about the possibility of direct applicability of our formulation 
(or that in~\citealp{EZFC}) in drawing quantitative conclusions in such situations, another problem arises.  
For  $m = 10^{-20} {\rm eV}$, the minimum wavelength $\lambda_{\rm min}$ of the 
fluctuations is actually more than an order of magnitude larger than  
the assumed initial cluster size. And as $\lambda_{\rm min} \sim  1/m$, for 
smaller masses $\lambda_{\rm min}$ is larger still.  
It is therefore unclear whether the fluctuations (including coherent core oscillations) would 
affect the internal structure of the cluster rather than  the  cluster as a whole. 

This places another limitation on obtaining strict constraints 
from the observed size of the central cluster of Eridanus II. 
At the same time, however, when the effective FDM granule  mass is larger than 
that of the cluster, one may expect energy equipartition 
between FDM quasiparticles and the cluster to result in significant motion of its centre of mass. 
As briefly discussed in the appendix, the displacement of the cluster form the centre of the galaxy, rather than its size, 
could in this case lead to constraint on $m$. However, a detailed 
examination of this issue  is beyond the scope of the present study.

\section{Conclusion}
\label{sec:conc}

In this study, we first extended  the model of~\cite{EZFC} to 
systems of classical point particles with spatial distributions that 
are statistically homogeneous, up to finite $N$ fluctuations. 
The associated power spectrum of density fluctuations is flat, 
corresponding to white noise. This case was not considered in~\cite{EZFC}, where 
we developed a model linking (random Gaussian) density fluctuations in a gaseous medium to potential 
fluctuations that can transform dark halo cusps into cores.  
In this case, the relevant dependence of the power spectrum on wave number 
is primarily of the form of a power law ($\mathcal{P}(k) \sim k^{-n}$, with $n  <  0$).

Application to white noise power spectrum  
leads to the standard two body relaxation time (if the maximal and 
minimal impact parameters are associated with maximal and minimal 
cutoffs in the flat power spectrum of density fluctuations; Section~\ref{sec:fixed}).  This is not all 
that surprising, as we evaluate the dynamical effects of finite-$N$ fluctuations on the motion of a test particle,
while ignoring their self gravity and the mean field (what we termed  the 'Jeans-Chandrasekhar Swindle'). These are the assumptions from which the usual two body relaxation time is derived, albeit by different means.

In~\cite{EZFC} we used the sweeping hypothesis, widely employed in turbulence theory, 
to transform spatial power spectra into the time domain, while assuming that 
all density fluctuation modes moved with the same velocity. Here we extend this by allowing 
each Fourier mode to be 'swept' with its own velocity and imposing mass 
conservation to determine its time evolution (Section~\ref{sec:dist}).  
If the mode velocities are uncorrelated with wavenumber 
(as expected of a homogeneous classical system), there arises a particularly simple 
relationship linking the spatio-temporal power spectrum to the spatial (equal time) 
power spectrum and the velocity distribution function (equation~\ref{eq:simp}). 
This then naturally leads to the full set of diffusion coefficients associated with  
standard relaxation theory (Appendix~\ref{app:diff}).

Next we consider systems of ultralight FDM axions, in the mean field limit and again assuming 
spatial homogeneity on large scales.  Given the aforementioned limit, the fluctuations here 
are not due to finite $N$ effects, but rather to the large de Broglie wavelength 
associated with the ultralight particles. The Jeans-Chandrasekhar swindle here 
requires that the de Broglie wavelength is small enough so that the self gravity of the fluctuations 
can be ignored --- that is much smaller than the Jeans length, which is of the order of the 
size of the physical system to be modelled. 
The sweeping hypothesis, generalized as described above, 
can still be used to obtain the spatio-temporal density power spectrum, and from this the force power spectrum 
and correlation function.  As in the case of gaseous fluctuations and classical 
point particle systems, the force correlation function can be inserted into a stochastic 
equation to evaluate the effect of relaxation due to the fluctuations and the associated timescale. 

There are two main differences with the classical case nevertheless. Although the relevant Schr\"odinger 
equation is that of a classical field (given the large number of bosons assumed to share the same states), 
the wave nature of the field naturally links the velocities at which Fourier density fluctuation modes can move 
to their wavenumber;  velocities and wavenumbers are therefore no longer uncorrelated, and 
mass conservation now requires that the wave function modes be swept at a
frequency-wavenumber dependent phase velocity.  
Furthermore, the usual quantum interference between wave function modes 
is present. The interference pattern,  arising from associating the density with the square of the wave function, 
leads to a power spectrum that is dependent on products of pairs of wavenumber (or equivalently velocity) distribution 
functions.  

In the diffusion limit, the effect of the FDM fluctuations on the motion on a classical test particle 
(namely in terms of increase in velocity dispersion), 
can still be written in the form familiar from standard two body relaxation theory.  
However the said interference effects imply that the resulting 
quasi-particles, through which the association with classical relaxation theory is made,
have effective mass and distribution function that involves integrals of the 
square of the velocity distribution function of the FDM axions. 
The resulting diffusion coefficients are equivalent to those derived in~\cite{BOFT}, though our approach involves
explicit evaluation of the force correlation function, with the diffusion limit taken as a final step. 

In the context of the present formulation, it becomes apparent that the procedure  
leading to the aforementioned effective quantities (and description in terms of quasiparticles),  
practically entails  taking the long wavelength limit. This is in turn associated with a white noise spatial power
spectrum of density fluctuations on scales larger than the de Broglie wavelength.  
In light of this approximation,  a Coulomb logarithm  arises; 
its choice determines the range of 
modes contributing to the white noise spectrum. 
This approach is particularly  well motivated when the distribution function is nearly constant for smaller 
velocities (larger wavelengths), 
before sharply cutting off beyond
a characteristic value (Section~\ref{sec:forcecorr_ax}; Appendix~\ref{app:Coul}).  
   
This is the case, for example, for a Maxwellian velocity distribution. 
We derive explicit expressions for the power spectrum and correlation function of FDM density fluctuations
and the mass fluctuations on different scales in this case (Section~\ref{sec:typical_dens}). 
These are of order one on scales of the order of the de Broglie wavelength and smaller,  
and then decay as Poisson noise on larger scales.  
As expected, the associated  
spectrum is flat (corresponding to white noise) on scales that are significantly larger 
than the de Broglie wavelength of the FDM axions, with a cutoff on smaller scales. 
This trend fits quite well the power spectrum from the numerical simulations of~\cite{Chan2018}.  
We also compare our results with correlation functions 
computed from numerical simulations by~\cite{Veltmaat2018}.
Our time correlation functions  nicely match theirs (especially when those are averaged 
over larger spherical regions). Their spatial correlation functions are well fit on smaller scales. 
They  then  however display a 
weakly declining tail at large scales that is not captured by our Gaussian decay. 
This could  be possibly related to residual correlations of 
the background density that remain after subtracting the radially averaged density 
field; for example, due to the non-sphericity of the halo and the finite size of the fluctuations, which 
also varies with radius due to changing de Broglie wavelength. 

We also estimate the effect of FDM fluctuations on galactic disks. If the 
total vertical velocity dispersion in the solar neighbourhood is attributed to such 
fluctuations, the FDM axion is constrained to have a mass $m \ga 0.3 \times 10^{-22} {\rm eV}$. 
Noting however the 
$\sim t^{1/3}$ growth of radial velocity dispersion with time implied by observations, further constraint 
can be  inferred. This is because the growth described by our diffusive model 
is necessarily $\sim t^{1/2}$, which contradicts the 
extensive recent study 
of~\cite{Mack_Bovy2019}. Reasoning that the FDM fluctuation 
contribution to the radial dispersion should therefore be limited to the errors in the aforementioned data, 
in order to avoid inconsistency, we find $m \ga 2 \times 10^{-22} {\rm eV}$
(Section~\ref{sec:verad}).  

This latter constraint  is similar those obtained by~\cite{AmLoeb2018} 
from evaluation of the effect of fluctuations on the dynamics of stellar streams, and also
those inferred from the dynamics of dark matter dominated objects such as dwarf and 'ultra-diffuse' galaxies  
(e.g.,~\citealp{Wasserm19}). 
It is however significantly weaker than those obtained by directly applying the model 
of~\cite{EZFC} to the central star cluster of the ultrafaint dwarf galaxy Eridanus II
(\citealp{Marsh2018}).  We applied our extended and generalized model to that situation, 
confirming that similar constraints can in principle be derived under the same assumptions, 
while discussing the limitations associated with these assumptions (Section~\ref{sec:comparison} and 
Appendix~\ref{App:Eri}).

The constraints from the disk dynamics derived here are also weaker than those inferred 
through entirely different means, such
as Lyman-$\alpha$ (\citealp{Lyman1_2017}; \citealp{BaldiLyman}) and 21 cm observations (\citealp{21cm1_2018}; \citealp{Hui_21cm}), and    
the environment around  supermassive black holes 
(e.g., \citealp{Superad}; \citealp{BarnBH19}; \citealp{NusserBH19}; \citealp{EllioMocz19}).   
While these methods are subject to their own uncertainties, our  model for the effect of FDM fluctuations
on disks, on the other hand, only takes 
into account the disk self gravity in the simplest possible way (by applying the vertical 
virial theorem; Section~\ref{sec:verad}). The disk self-gravitating response may in principle modify 
the effect of fluctuations. If it is similar in nature to the response of the nonradial modes in the perturbed haloes 
studied in~\cite{EZFC}, it may in fact significantly amplify and enhance the effect of the imposed stochastic fluctuations. 
This should be worth studying in detail in future work, as in a disk such effects may in fact be expected to be 
even more prominent.



\section*{Acknowledgements}
We thank Jens Niemeyer and Jan Veltmaat for very helpful communications and 
for sharing the spatial correlation function of 
simulated FDM haloes, and Scott Tremaine for valuable comments.  
This project was supported financially by the Science
and Technology Development Fund (STDF), Egypt. Grant
No. 25859, and by the Franco-Egyptian Partenariat Hubert Curien (PHC)
Imhotep  project 2019 42088ZK.




\bibliographystyle{mnras}
\bibliography{FDM_Flucs_vfez_fin_A} 




\appendix


\section{Velocity dispersion for a white noise power spectrum}
\label{app:2body}

Adapting the theoretical framework introduced in \cite{EZFC} to a test particle affected by stationary stochastic density fluctuations with a time-dependent white noise power spectrum, resulting from all modes move with the same velocity $v_r$ relative to the test particle, yields a velocity dispersion after time $T$ given by Eq.~(\ref{eq:detdisp}). This equation is the sum of 
\begin{equation}
\langle (\Delta v_p)^2 \rangle_1 = \frac{2DT}{v_r} \int_0^T \frac{{\rm Si}(k_x v_r t)- {\rm Si}(k_m v_r t) }{t} dt
\end{equation}
and
\begin{equation}
\langle (\Delta v_p)^2 \rangle_2 = - \frac{2D}{v_r} \int_0^T \left[ {\rm Si}(k_x v_r t)- {\rm Si}(k_m v_r t) \right] dt, 
\end{equation}
where $2\pi/k_m$ and $2\pi/k_x$ are the maximum and minimum cutoff scales of the  power spectrum and Si denotes the Sine integral function. 
These two terms can be rewritten as 
\begin{equation}
\langle (\Delta v_p)^2 \rangle_1 = \frac{2DT}{v_r} \left[ \int_0^{u_x} \frac{{\rm Si}(u)}{u} du -  \int_0^{u_m} \frac{{\rm Si}(u)}{u} du\right]
\end{equation}
and
\begin{equation}
\langle (\Delta v_p)^2 \rangle_2 = \frac{2DT}{v_r} \left[\frac{1}{u_m}\int_0^{u_m} {\rm Si}(u) du -  \frac{1}{u_x}\int_0^{u_x} {\rm Si}(u) du \right]
\end{equation}
with $u_x=k_x v_r T$ and $u_m=k_m v_r T$. 

In the diffusion limit, where both $u_x,u_m\gg 1$, we can introduce an intermediate $u_d<\min(u_x,u_m)$ such that $u_d \gg 1$ and write 
\begin{equation}
\label{eq:DV1_2}
\langle (\Delta v_p)^2 \rangle_1 = \frac{2DT}{v_r} \left[ \int_{u_d}^{u_x} \frac{{\rm Si}(u)}{u} du -  \int_{u_d}^{u_m} \frac{{\rm Si}(u)}{u} du\right].
\end{equation}
Since ${\rm Si}(u) = \pi/2 -\cos(u)/u+ \mathcal{O}(1/u^2)$ when $u\gg1$, this equation leads to 
\begin{equation}
\langle (\Delta v_p)^2 \rangle_1 = \frac{\pi D T}{v_r} \left[ {\rm ln}\left(\frac{u_x}{u_m}\right) +\mathcal{O}\left( \frac{1}{u_m}\right)\right]
\end{equation}
when $u_x,u_m\gg 1$. 
Since $\int_0^u {\rm Si}(t) dt = u {\rm Si}(u) + {\rm cos}(u)-1$, $\int_0^u {\rm Si}(u) du = u \pi/2 -1 +\mathcal{O}(1/u^2)$ when $u\gg 1$ and it yields
\begin{equation}
\langle (\Delta v_p)^2 \rangle_2 = \frac{2DT}{v_r} \left[ \frac{1}{u_x}-\frac{1}{u_m} + \mathcal{O}\left( \frac{1}{u_m^2}\right) \right]. 
\end{equation}
Hence in the diffusion limit, 
\begin{equation}
\langle (\Delta v_p)^2 \rangle = \frac{\pi D T}{v_r} \left[ {\rm ln}\left(\frac{u_x}{u_m}\right) +\mathcal{O}\left( \frac{1}{u_m}\right)\right],
\end{equation}
which leads to Eq.~(\ref{eq:tbs}). 


\section{Diffusion coefficients}
\label{app:diff}

In general, the relation between the force and potential Fourier components 
can be written as 
\begin{equation}
F_{\bf k} = - i {\bf k} \phi_{\bf k}.
\end{equation}
Equation~(\ref{eq:pfk}) hence generalizes, for two modes ${\bf k_i}$ and $\bf k_j$, to 
\begin{equation}
\mathcal{P}_F (k_i, k_j, t) = (4 \pi G \rho_0)^2 \mathcal{P} ({\bf k}, t) \frac{k_i k_j}{k^4}
\end{equation}
and Eq.  ({\ref{eq:FCF}) becomes
	\begin{equation}
	\langle F_i (0, 0) F_j (r, t) \rangle = \frac{(4 \pi G \rho_0)^2}{(2 \pi)^3} \int \frac{k_i k_j}{k^4} e^{i {\bf k}. {\bf r}} \mathcal{P} ({\bf k}, t) d {\bf k}.
	\end{equation}
	Using equation (\ref{eq:2b_ps}) this can be written as
	\begin{equation}
	\langle F_i (0, 0) F_j (r, t) \rangle = \frac{2}{\pi} m G^2  \int f ({\bf v}) d {\bf v} \int  \frac{k_i k_j}{k^4}
	e^{i {\bf k} . ({\bf v}_p - {\bf v}) t} d {\bf k}
	\end{equation}
	with $\vec{r}=\vec{v}_p t$ as in Eq.~(\ref{eq:F00Frt}).
	Through straightforward generalization of equation~(\ref{eq:corrsp}), the non-transient term describing the
	growth of  the velocity dispersion under the action of  the density fluctuations can be expressed as 
	\begin{equation}
\begin{split}
	\langle \Delta v_{p i} \Delta v_{p j} \rangle &=\\ &\frac{4}{\pi} m G^2  T \! \int_0^T \!\!\left(\int \! f ({\bf v}) d {\bf v} \int \! \frac{k_i k_j}{k^4}
	e^{i{\bf k} . ({\bf v}_p - {\bf v}) t } d {\bf k} \! \right) \! d t.
\end{split}
	\label{eq:DviDvj}
	\end{equation} 
	Taking spherical coordinates with z-axis along vector \mbox{${\bf v}_p - {\bf v}= {\bf V}_0 = - {\bf v}_r$} yields 
	\begin{equation}
	\int \frac{k_i k_j}{k^4} e^{i{\bf k} . ({\bf v}_p - {\bf v}) t } d {\bf k}
	= 
	\int \frac{k_i k_j}{k^2}
	e^ {i k  V_0 t \cos\theta }  \sin\theta d \theta d \phi dk, 
	\end{equation}
	%
	%
	where $V_0 = |{\bf V}_0| = |{\bf v}_r|$.
	Taking the large $T$ (diffusion limit) in equation~(\ref{eq:DviDvj}), the integration of the exponential 
over time involves a delta function with 
	$\cos \theta$ as argument, meaning that only wave number vectors normal to ${\bf V}_0$ contribute. This results in 
	\begin{equation}
	\langle \Delta v_{p i} \Delta v_{p j} \rangle = 4  G^2  m T \int \frac{f({\bf v})}{V_0} d {\bf v} \int  \frac{k_i k_j}{k^3} d k  d \phi. 
	\label{eq:prediff}
	\end{equation}
	Furthermore, since $\cos \theta = 0$ for contributing wave number vectors, 
	$k_x = k \cos \phi$, $k_y = k \sin \phi$ and $k_z = 0$. The components along some general 
	unit vectors $\hat{\bf e}_i$ and $\hat{\bf e}_i$ are: 
	$k_i = k_x (\hat{\bf e}_x . \hat{\bf e}_i ) + k_y  (\hat{\bf e}_y . \hat{\bf e}_i)$ 
	and $k_j =  k_x (\hat{\bf e}_x . \hat{\bf e}_j ) + k_y  (\hat{\bf e}_y . \hat{\bf e}_j)$. Thus  
	we can integrate equation~({\ref{eq:prediff}) over $\phi$ and $k$ to get   
		\begin{align}
		\nonumber
		\langle \Delta v_{p i} \Delta v_{p j}\rangle =  & ~4 \pi m  G^2  T 
		\ln \frac{k_x}{k_m} \times\\
		&\int \frac{f ({\bf v})}{V_0}  [(\hat{\bf e}_x . \hat{\bf e}_i )
		(\hat{\bf e}_x . \hat{\bf e}_j ) +  
		(\hat{\bf e}_y . \hat{\bf e}_i)   (\hat{\bf e}_y . \hat{\bf e}_j )] 
		d {\bf v}. 
		\end{align}
		As, in the coordinate system where the integration was performed, we took
		the z-axis along ${\bf V}_0$, we have \mbox{$(\hat{\bf e}_z . \hat{\bf e}_i ) = V_{0 i}/V_0$}. 
		In addition
		\begin{equation}
		(\hat{\bf e}_x . \hat{\bf e}_i )  (\hat{\bf e}_x . \hat{\bf e}_j )  +  (\hat{\bf e}_y . \hat{\bf e}_i )  (\hat{\bf e}_y . \hat{\bf e}_j )
		+   (\hat{\bf e}_z . \hat{\bf e}_i )  (\hat{\bf e}_z . \hat{\bf e}_j ) \! = \! (\hat{\bf e}_i . \hat{\bf e}_j ) \! = \! \delta_{i j}.
		\end{equation}
		Therefore, 
		\begin{equation} 
		\langle \Delta v_{p i} \Delta v_{p j}\rangle =  4 \pi m G^2  T 
		\ln \frac{k_x}{k_m}
		\int \frac{f ({\bf v})}{V_0} \left[\delta_{i j} - \frac{V_{0 i} V_{0 j}}{V_0^2}\right] d {\bf v}. 
		\end{equation}

		As this final form of the velocity dispersions  resulting from fluctuating force already incorporates the diffusion limit, the   second order diffusion coefficients $D [\Delta v_{i} \Delta v_{j}]$
are simply given by~\footnote{We drop 
the subscript $p$ in the square brackets of the diffusion coefficients to simplify (and connect to standard) notation.} 
		\begin{equation}
		D [\Delta v_{i} \Delta v_{j}] = \frac{\langle \Delta v_{p i} \Delta v_{p j}\rangle}{T},
		\end{equation}
    which leads to the same standard form for the diffusion coefficients (as in Binney \& Tremaine 2008, Appendix L)
when   $\ln \frac{k_x}{k_m}$ is idenstified with the Coulomb logarithm $\ln \Lambda$. 
		A test particle moving in a fluctuating field that is stationary and random Gaussian 
		(fully defined by two point correlation function and power spectrum) 
		will also experience a drag force that is related to the correlation function of the force 
		fluctuations (e.g. \citealp{Kubo1966, BT}). This being the case, the first order diffusion coefficients will be related to the 
		second order 
		ones by the fluctuation dissipation relations 
		\begin{equation}
		D [\Delta v_{i}] = \frac{1}{2} \sum_j \frac{\partial }{\partial v_{p j}} D [\Delta v_{i} \Delta v_{j}]. 
		\end{equation}
		
		For isotropic velocities --- i.e. in systems with  distribution functions depending on $v$
		rather than ${\bf v}$ --- the Cartesian diffusion coefficients are related to  the 
		coefficients in the directions parallel and perpendicular to a test particle's motion by
		(Binney \& Tremaine 2008, Appendix L)
		\begin{equation}
		D [\Delta v_i] = \frac{v_{p i}}{v_p} D [\Delta v_\parallel]
		\end{equation}
		and
		\begin{align}
       \nonumber
		D   [\Delta v_{i} \Delta v_{j}]  =\frac{v_{p i} v_{p j}}{v_p^2} \left(D [(\Delta v_\parallel)^2]\right. & - \left.\frac{1}{2} D [(\Delta v_\perp)^2] \right) \\
		& +\frac{1}{2} \delta_{i j} D [(\Delta v_\perp)^2]. 
		\end{align}
		For a Maxwellian velocity distribution with one dimensional dispersion $\sigma$
		(Eq.~\ref{eq:Max})
		\begin{equation}
		D [\Delta v_\parallel] = - \frac{4 \pi G^2 \rho_0 m \ln \Lambda}{\sigma^2} {\rm G} (X),
		\end{equation}
		\begin{equation}
		D [(\Delta v_\parallel)^2] = \frac{\sqrt{2}~4 \pi G^2 \rho_0 m \ln \Lambda}{\sigma}  \frac{{\rm G} (X)}{X},
		\end{equation}
		and
		\begin{equation}
		D [(\Delta v_\perp)^2] = \frac{\sqrt{2}~4 \pi G^2 \rho_0 m \ln \Lambda}{\sigma} \left[\frac{{\rm erf}  (X) - {\rm G} (X)}{X}  \right].
		\end{equation}
		Here $X = v_p/\sqrt{2} \sigma$ and 
		\begin{equation}
		{\rm G} (X) = \frac{1}{2 X^2} \left[{\rm erf}  (X) - \frac{2 X}{\sqrt{\pi}} e^{-X^2} \right].
		\end{equation}
		
		In the case of FDM axions, $m$ is replaced by the effective mass $m_{\rm eff}$ and the distribution
		function $f$ is replaced by $f_{\rm eff}$ as given by  equations (\ref{eq:meff}) and~(\ref{eq:feff}). 
		Furthermore, when $f$ is Maxwellian,   the effective distribution function is also a Maxwellian 
		with $\sigma_{\rm eff} = \sigma/\sqrt{2}$. 
		Thus in the FDM case, the above relations for 
		the parallel and perpendicular diffusion coefficients are still valid once $m$ is replaced with $m_{\rm eff}$, 
		$\sigma$ with $\sigma_{\rm eff}$ and $X$ with $X_{\rm eff} = v_p/(\sqrt{2} \sigma_{\rm eff}) = v_p/ \sigma$.  

As noted by~\cite{BOFT}, 
the  diffusion coefficients, thus defined, do not tend to the classical 
point particle limit as $m_{\rm eff} \rightarrow 0$.  This is because, as
mentioned (while introducing equations~\ref{eq:schrod} and~\ref{eq:PoiV}),
the mean field limit is already implied from the start by the description in terms of a classical 
field  arising from large occupation numbers (i.e., large number of particles existing in the same state).  
The classical collisionless limit, with diffusion coefficients vanishing, is therefore naturally 
arrived at as $m_{\rm eff} \rightarrow 0$, as fluctuations due to finite de Broglie wavelength 
vanish. The classical point particle limit can be recovered by considering  wave packets representing 
individual particles~(\citealp{BOFT}, Appendix~A).


\section{Power spectrum of a free axion system}
\label{app:BOFT}

Ignoring the self-gravity term in Eq.~(\ref{eq:schrod}) leads to an axion wavefunction that can be written as 
\begin{equation}
\psi({\bf r},t) = \int \phi_{\bf k} e^{i{\bf k}.{\bf r}-i\omega({\bf k})t}  d{\bf k}
\end{equation}
with 
\begin{equation}
\omega({\bf k}) = \frac{\hbar k^2}{2m} 
\end{equation}
as indicated in Section~\ref{sec:axion_fluc}. The assumption of a free field is justified there in terms of the 'Jeans-Chandrasekhar swindle'; 
effectively the assumption of an  infinite medium that is statistically homogeneous on large scales, with the only contributions 
to the potential affecting a test particle coming from fluctuations around a mean field that is subtracted away.  In the case of a FDM 
field this implies that the characteristic fluctuation scale (roughly the de Broglie wavelength) 
is small compared to the Jeans length --- 
effectively the size of an actual inhomogeneous self-gravitating system  ---  
since the self gravity of the fluctuations is ignored. 
 
We assume that the ensemble averages of $\phi_{\bf k}$ satisfy $\langle \phi_{\bf k} \rangle = 0$ and \mbox{$\langle \phi_{\bf k}\phi^\star_{\bf k^\prime}\rangle = f_\vec{k}({\bf k}) \delta_{\rm D}({\bf k}-{\bf k^\prime})$} where $\delta_{\rm D}$ is the Dirac delta function, i.e., $\langle \phi_{\bf k} \phi^\star_{\bf k^\prime}\rangle = 0$ for ${\bf k}\neq{\bf k^\prime}$ and the mean axion density $\rho_0=\langle |\psi({\bf r},t)|^2 \rangle = \int f_\vec{k}({\bf k}) d{\bf k}$. 
They are therefore modes of a complex Gaussian random field.   
This is consistent with the assumption, as in \cite{EZFC},  that the 
fluctuations giving rise to the stochastic dynamics 
describe a statistically homogeneous Gaussian random field, completely characterized 
by its power spectrum and correlation function. 

Since the functions $\phi_{\bf k}$ are complex valued Gaussian random variables, 
 Isserlis' (or Wick's) theorem applies and 
\begin{multline}
\label{eq:axion_ensembles}
\langle \phi_{\bf k_1} \phi^\star_{\bf k_2} \phi_{\bf k_3} \phi^\star_{\bf k_4}\rangle = 
f_\vec{k}({\bf k_1}) f_\vec{k}({\bf k_3}) \delta_{\rm D}({\bf k_1}-{\bf k_2}) \delta_{\rm D}({\bf k_3}-{\bf k_4})\\
+
f_\vec{k}({\bf k_1}) f_\vec{k}({\bf k_2}) \delta_{\rm D}({\bf k_1}-{\bf k_4}) \delta_{\rm D}({\bf k_2}-{\bf k_3})
\end{multline}

Given these assumptions, the axion density fluctuations $\delta({\bf r},t) = |\psi({\bf r},t)|^2/\rho_0 - 1 $ arising from the Schr\"odinger-Poisson system are described by the density contrast
\begin{equation}
\delta({\bf r},t) 
= \frac{1}{\rho_0}\iint \phi_{\bf k} \phi^\star_{\bf k^\prime}e^{i[{\bf k}-{\bf k^\prime}].{\bf r}-i[\omega({\bf k})-\omega({\bf k^\prime})] t} d{\bf k} d{\bf k^\prime} -1, 
\end{equation}
whose two-point correlation function $ C ({\bf r},t) = \langle \delta(0,0)\delta({\bf r},t)\rangle$ is 
\begin{equation}
\label{eq:axion_xi}
C ({\bf r},t) 
= \frac{1}{\rho_0^2} \iint f_\vec{k}({\bf k}) f_\vec{k}({\bf k^\prime}) e^{i[{\bf k}-{\bf k^\prime}].{\bf r}-i[\omega({\bf k})-\omega({\bf k^\prime})]t} d{\bf k}d{\bf k^\prime}
\end{equation}
using Eq.\ref{eq:axion_ensembles}. 
Taking the Fourier transform of Eq.~\ref{eq:axion_xi} yields the power spectrum
\begin{align}
\nonumber
\mathcal{P}(&{\bf k},t) =
\frac{(2\pi)^3}{\rho_0^2} \times \\  
&\iint f_\vec{k}({\bf k_1}) f_\vec{k}({\bf k_2}) e^{-i[\omega({\bf k_1})-\omega({\bf k_2})]t} \delta_{\rm D} (\!{\bf k}\!-\!{\bf k_1}\!+\!{\bf k_2}\!) d{\bf k_1}d{\bf k_2}. 
\end{align}


By analogy with the case where each mode of the density perturbation is swept by a given velocity, we can introduce the velocities 
\begin{equation}
{\bf v}_i = \frac{\hbar {\bf k}_i}{m}, ~ {\bf v}_c=\frac{{\bf v}_1+{\bf v}_2}{2}, ~\mathrm{and}~ {\bf v}_d=\frac{{\bf v}_1-{\bf v}_2}{2}
\end{equation}
such that 
\begin{align}
\nonumber
\mathcal{P}(&{\bf k},t) =
\frac{(2\pi)^3}{\rho_0^2}\times\\
&\iint f({\bf v_1}) f({\bf v_2}) e^{-im_\hbar {\bf v}_c.{\bf v}_d t} \delta_{\rm D} ({\bf k}-m_\hbar {\bf v}_d) d{\bf v_1}d{\bf v_2}
\end{align}
with $m_\hbar=2m/\hbar$ and $f_{\vec{k}}({\bf k}_i) d{\bf k}_i= f ({\bf v}_i) d{\bf v}_i$, i.e., \mbox{$f({\bf v}_i) = f_\vec{k}(m{\bf v}_i/\hbar) \times m/\hbar$}. 
%


\section{Axion fluctuations with a Maxwellian velocity distribution}
\label{app:max}

\subsection{Equal time power spectrum}

The equal time power spectrum of the density contrast of a free axion system is given by Eq.~(\ref{eq:axion_onet}). For a Maxwellian velocity distribution (Eq.~\ref{eq:Max}), it yields: 
\begin{align}
\nonumber
\mathcal{P}(\vec{k},0)
& =\frac{8}{m_\hbar^2 \sigma^6}
\int e^{-\frac{1}{2\sigma^2} \left[ (\vec{v}+\vec{k}/m_\hbar )^2+(\vec{v}-\vec{k}/m_\hbar )^2\right]} d\vec{v}\\
\nonumber
& = \frac{8}{m_\hbar^2 \sigma^6} e^{-\frac{k^2}{m_\hbar^2 \sigma^2}} 
\int e^{-\frac{\vec{v}^2}{\sigma^2}}d\vec{v}\\
\mathcal{P}(\vec{k},0) 
& = \left( \frac{2\sqrt{\pi}}{m_\hbar \sigma}\right)^{3} e^{-\frac{k^2}{m_\hbar^2 \sigma^2}}. 
\end{align}

\subsection{Correlation function of the density contrast}

The power spectrum of the density contrast of a free axion system can more generally be expressed through equation~\ref{eq:PS_Ax} as
\begin{equation}
\mathcal{P}(\vec{k},t) = \frac{(4\pi)^3}{m_\hbar^3\rho_0^2} 
\int f(\vec{v}+\vec{k}/m_\hbar) f(\vec{v}-\vec{k}/m_\hbar) e^{-i\vec{v}.\vec{k}t} d\vec{v}. 
\end{equation}
The associated correlation function is the inverse Fourier transform of this power spectrum, 
\begin{align}
\nonumber
\langle \delta(0,0)&\delta(\vec{r},t)\rangle
= \frac{1}{(2\pi)^3} \int \mathcal{P}(\vec{k},t) e^{i\vec{k}.\vec{r}} d\vec{k}\\
\nonumber
& = \frac{1}{(\pi \sigma^2 m_\hbar)^3} 
\iint e^{-\frac{1}{\sigma^2}( \vec{v}^2+\vec{k}^2/m_\hbar^2)+i(\vec{r}-\vec{v}t).\vec{k}} d\vec{v} d\vec{k}\\
\nonumber
& = \frac{1}{(\pi \sigma^2 m_\hbar)^3} 
\int e^{-\frac{\vec{v}^2}{\sigma^2}} \left( \int e^{-\frac{\vec{k}^2}{\sigma^2m_\hbar^2}+ i(\vec{r}-\vec{v}t).\vec{k} } d\vec{k}\right) d\vec{v}\\
\nonumber
& = \frac{1}{(\pi\sigma^2)^{3/2}} \int e^{-\frac{\vec{v}^2}{\sigma^2} - \frac{m_\hbar^2\sigma^2(\vec{r}-\vec{v}t)^2}{4}} d\vec{v}\\
&= \frac{1}{(\pi\sigma^2)^{3/2}} \int e^{-\frac{\vec{v}^2}{\sigma^2} - \frac{(\vec{r}-\vec{v}t)^2}{\lambda_\sigma^2}} d\vec{v}, 
\end{align}
with $\lambda_\sigma =2/m_\hbar\sigma = \hbar/m\sigma$ an associated wavelength
(equal to $1/2 \pi$ the de Broglie wavelength connected to motion at speed $\sigma$).   
The exponent in the exponential can be rewritten
\begin{align}
\nonumber
&-\frac{\vec{v}^2}{\sigma^2} - \frac{(\vec{r}-\vec{v}t)^2}{\lambda_\sigma^2}
=\\
&-\frac{1+t^2\sigma^2/\lambda^2}{\sigma^2}\left[\left( \vec{v}-\frac{t\sigma^2/\lambda_\sigma^2}{1+t^2\sigma^2/\lambda_\sigma^2} \vec{r}\right)^2 +\frac{\vec{r}^2 \sigma^2/\lambda_\sigma^2}{(1+t^2\sigma^2/\lambda_\sigma^2)^2}  \right]
\end{align}
such that
\begin{align}
\langle \delta(0,0)\delta(\vec{r},t)\rangle
& =
\frac{1}{(\pi\sigma^2)^{3/2}} e^{-\frac{r^2/\lambda_\sigma^2}{1+t^2\sigma^2/\lambda_\sigma^2}}
\int e^{-\frac{1+t^2\sigma^2/\lambda_\sigma^2}{\sigma^2}\vec{v^\prime}^2}d\vec{v^\prime}
\end{align}
with the change of variable $\vec{v^\prime}=\vec{v}-\vec{r}t\sigma^2/(\lambda_\sigma^2+t^2\sigma^2)$. We finally have 
\begin{align}
\langle \delta(0,0)\delta(\vec{r},t)\rangle =
\frac{1}{(1+t^2\sigma^2/\lambda_\sigma^2)^{3/2}} e^{-\frac{r^2/\lambda_\sigma^2}{1+t^2\sigma^2/\lambda_\sigma^2}}. 
\end{align}

\section{Coulomb logarithm for FDM axion systems}
\label{app:Coul}

We would like to estimate the value of the argument of the Coulomb logarithm for FDM axion systems, 
as defined by equation~(\ref{eq:LFDM}).

In this regard, we  first 
note that the ratio $v_{dx}/v_{dm}$ 
is associated with a minimal and maximal scale ($\lambda_{\rm min}$ and $\lambda_{\rm max}$) 
through the de Broglie wavelength.  We therefore relate  it to the  
maximal and minimal velocities associated with these wavelength, such that
the ratio of the velocities is related to that of the minimal and maximal wavelength 
as   
$v_{dx}/v_{dm} = v (\lambda_{\rm min})  / v (\lambda_{\rm max})  = \lambda_{\rm max} / \lambda_{\rm min}$.   

Second, we note that the Coulomb logarithm appeared in our calculation 
with the approximation that lead 
from equation (\ref{eq:fullqcorr}) to 
(\ref{eq:axforccorr}).  Strictly speaking, the evaluation 
of~(\ref{eq:fullqcorr}) should involve the full integration over 
a factor $1/v$ multiplied by the  phase space distribution functions $f (v_1) f(v_2)$. For the Maxwellian 
distributions adopted here, this entails a sharp cutoff in the integrand 
at speeds $v > \sigma$. Physically, this would also be 
motivated by the fact that  it approximately corresponds to the length scale associated with the 
effective mass.  On the other hand, 
no such cutoff exists at small speeds, so the maximum wavelength entering
into  Coulomb logarithm can be extended up to the range of validity of our formulation, as we discuss in specific cases below.

\subsection{Singular isothermal sphere and effect on disk}

In light of the comments above, 
the minimal scale should be of the order of that determined by the effective mass scale.
Thus, setting $m_{\rm eff} = \frac{4}{3} \pi \rho_0 (\lambda_{\rm min})^3$, 
and using equation (\ref{eq:meffM}) to evaluate $m_{\rm eff}$ for a Maxwellian distribution, we estimate
\begin{equation}
\lambda_{\rm min} =  \left(\frac{3}{4}\right)^{1/3} \pi^{1/6} \frac{\hbar}{m \sigma},
\label{eq:lmin}
\end{equation}
and  
\begin{equation}
v (\lambda_{\rm min}) = \frac{h}{m \lambda_{\rm min}} = \left(\frac{4}{3}\right)^{1/3} \sqrt{2} \pi^{5/6} v_{\rm circ},
\end{equation}
where $v_{\rm circ} = \sqrt{2} \sigma$. 

The description in terms of  fluctuations of    
Gaussian random fluctuations that affect the particle velocities
locally (and associated diffusion limit), 
would not apply if  the characteristic timescale of fluctuations is smaller than the 
natural timescales associated with the motion of the perturbed (test) particle. In this 
case the fluctuating potential changes slowly along the particle trajectory, affecting 
it non-locally and adiabatically, rather than in as stochastic dynamical process 
with a diffusion limit.~\footnote{Longer wavelengths can still induce bending modes 
that may  dissipate their energy into heating a disk.  Whether such effects are important, and whether their scaling 
mimics those of diffusive processes derived here (e.g., equation \ref{eq:sigZ}), 
may be considered in future simulations  examining the self consistent response of 
a gravitating disk to induced FDM fluctuations, including the spectrum of
response frequencies that may be thus excited.} 
 
The minimal velocity can in this context be connected to the maximum scale of fluctuations 
that do not violate this assumption. The associated condition is 
\begin{equation}
\frac{T (\lambda_{\rm max})}{T_p} = 1
\label{eq:TT}
\end{equation}
where $T (\lambda_{\rm max}$) is the timescale related to the maximal wavelength
$T (\lambda_{\rm max}) = \lambda_{\rm max}/ v (\lambda_{\rm max})$ and $T_p$ is the characteristic 
period of radial and vertical oscillations of the test particle. In the epicyclic  
approximation, for nearly circular trajectories, this latter timescale is of the order of the circular 
orbit period (e.g., Binney \& Tremaine 2008). Given the ultimate weak logarithmic, 
for a test particle orbit of radius $R$ (e.g., representing a disk star)
we therefore just set $T_p =  v_{\rm circ}/ 2 \pi R$. 
Inserting these values into (\ref{eq:TT}),  with $v (\lambda_{\rm min}) = h/m \lambda_{\rm max}$,
implies  that 
\begin{equation}
v^2 (\lambda_{\rm max}) = \frac{\lambda (v_{\rm circ})}{2 \pi R} v^2_{\rm circ},
\label{eq:vlmax}
\end{equation}
where $\lambda (v_{\rm circ}) = h/m v_{\rm circ}$.

In accordance with our association of the argument of the Coulomb logarithm with
the ratio of maximal and minimal velocities we thus have 
\begin{equation}
\Lambda = 
\frac{v (\lambda_{\rm min})}{v (\lambda_{\rm max})} 
= \left(\frac{4}{3}\right)^{1/3} \sqrt{2}~\pi^{5/6}~\left(\frac{2 \pi R}{\lambda (v_{\rm circ})}\right)^{1/2}.
\label{eq:Lambda}
\end{equation}
For typical solar neighborhood parameters, 
\begin{equation}
\frac{\lambda (v_{\rm circ})}{2 \pi} = \frac{\hbar}{m v_{\rm circ}}  =
0.096~{\rm kpc} ~\left(\frac{10^{-22} {\rm eV}}{m}~\frac{200 {\rm km/s}}{v_{\rm circ}}\right), 
\end{equation}
and 
\begin{equation}
\Lambda = 36.9~\left(\frac{r}{8 {\rm kpc}}~ \frac{m}{10^{-22} {\rm eV}}~\frac{v_{\rm circ}}{200 {\rm km/s}}\right)^{1/2}. 
\end{equation}

This estimate takes into account the quantum origin of the fluctuations, which imposes a relation between the wavelength and frequency of density fluctuation that are fundamentally associated with 
an interference pattern between the wave function modes $\phi_{\bf k}$ (as detailed in Appendix~\ref{app:BOFT} above). 
On the other hand,  if the FDM fluctuations are considered to correspond to motions of classical particles of 
mass $m_{\rm eff}$ and  'size' of order $\lambda_{\rm min}$, the Coulomb logarithm argument $\Lambda$ is expected to be significantly 
larger than derived above. For, in this case
$\Lambda = b_{\rm max}/ b_{\rm min}$, 
with $b_{\rm min} \approx \lambda_{\rm min}$.  And imposing condition (\ref{eq:TT}) leads to 
$2 \pi R/ v_{\rm circ} \approx 2 b_{\rm max}/ v_r$, where $v_r \approx v_{\rm circ}$ is the relative velocity of the 
field particle perturbing the test particle at encounter parameter $b_{\rm max}$.
So $b_{\rm max}/  b_{\rm min} \approx \pi R/\lambda_{\rm min}$. As $\lambda_{\rm min}$ is of the same order 
of $\lambda (v_{\rm circ})$,  then $\Lambda \approx R/\lambda (v_{\rm circ})$, which is significantly larger than 
$\Lambda \sim \sqrt{R/\lambda (v_{\rm circ})}$ 
derived above (equation \ref{eq:Lambda}). 

\subsection{The case of Eridanus II}

In this case we take $T_p$ to correspond to the Keplerian  period;
$T_p = 2 \pi R/v_p$, with $v_p^2 = G M_c/R$ and  
 $M_c = 2000 {\rm M_\odot}$ the central cluster's mass (for observationally measured parameters of Eridamus II
and its central cluster see~\citealp{EriobsI,EriobsII}).  
Assuming an isothermal core for the dwarf galaxy, 
Eq.~(\ref{eq:lmin}) is still valid, with $\sigma = 6.9 {\rm km/s}$ corresponding to the
measured value.  Condition~(\ref{eq:TT}) now leads to
\begin{equation} 
v^2 (\lambda_{\rm max}) = \frac{\lambda (v_p)}{2 \pi R} v^2_p,
\end{equation}
and therefore
\begin{equation} 
\Lambda = 
\frac{v (\lambda_{\rm min})}{v (\lambda_{\rm max})} 
= 2~\left(\frac{4}{3}\right)^{1/3}~\pi^{5/6}~\left(\frac{2 \pi R}{\lambda (v_p)}\right)^{1/2} \frac{\sigma}{v_p}.
\label{eq:Lambdac}
\end{equation}
For  $M_c = 2000 {\rm M_\odot}$, $v_p = 0.83~{\rm km/s} \sqrt{{\rm 13 pc}/R}$, so that  
\begin{equation}
\frac{\lambda (v_p)}{2 \pi} = \frac{\hbar}{m v_p}  =
2.3~{\rm kpc} ~\left(\frac{10^{-21} {\rm eV}}{m}~\sqrt{\frac{R}{13 {\rm pc}}}\right), 
\end{equation}
and
\begin{equation}
\Lambda = 3.57~ \left(\frac{R}{13 {\rm pc}}\right)^{3/4} \left(\frac{m}{10^{-21} {\rm eV}}\right)^{1/2}.
\label{eq:CouLEri}
\end{equation}

Clearly $\Lambda$  must be larger than unity for the formulation to make any sense. Thus 
for an initial cluster size of $2 ~{\rm pc}$, we must have $m > 1.3 \times 10^{-21} {\rm eV}$. 
This is similar to condition (5) of~\cite{Marsh2018}. 
It reflects the necessity that the minimal-wavelength modes considered
do not affect the cluster particles non-locally and adiabatically, rather than 
be characterized by a random Gaussian process with associated diffusion limit
as assumed here. 

Note that we require  $k_m |{\bf v}_p - {\bf v}| t \gg 1$ for all velocities ${\bf v}$, rather 
than simply $k_x |{\bf v}_p - {\bf v}| t \gg 1$  as assumed in~\cite{Marsh2018}. Both 
conditions are satisfied for a cluster age $t = 3 ~{\rm Gyr}$; which eliminates the time 
dependence of the Coulomb logarithm in~\cite{Marsh2018}, resulting from the 
assumption~$k_m |{\bf v}_p - {\bf v}| t \ll 1$. Finally, note that a maximal 
wavelength cutoff on the basis of guaranteeing locality and non-adiabaticity of fluctuations
may not be necessary when the cluster size is much smaller than the wavelength considered, 
as in this case one can consider its stars stationary and locality is guaranteed. However, 
this case raises questions as to whether one should consider the cluster as a whole, rather than 
individual stars,  to be affected by the fluctuations. We 
further comment on this in the next appendix.

\section{Revisiting Eridanus II}
\label{App:Eri}

\begin{figure}
	\centering
	\includegraphics[width=1 \columnwidth,trim={0.cm 0.1cm 1.2cm 1cm},clip]{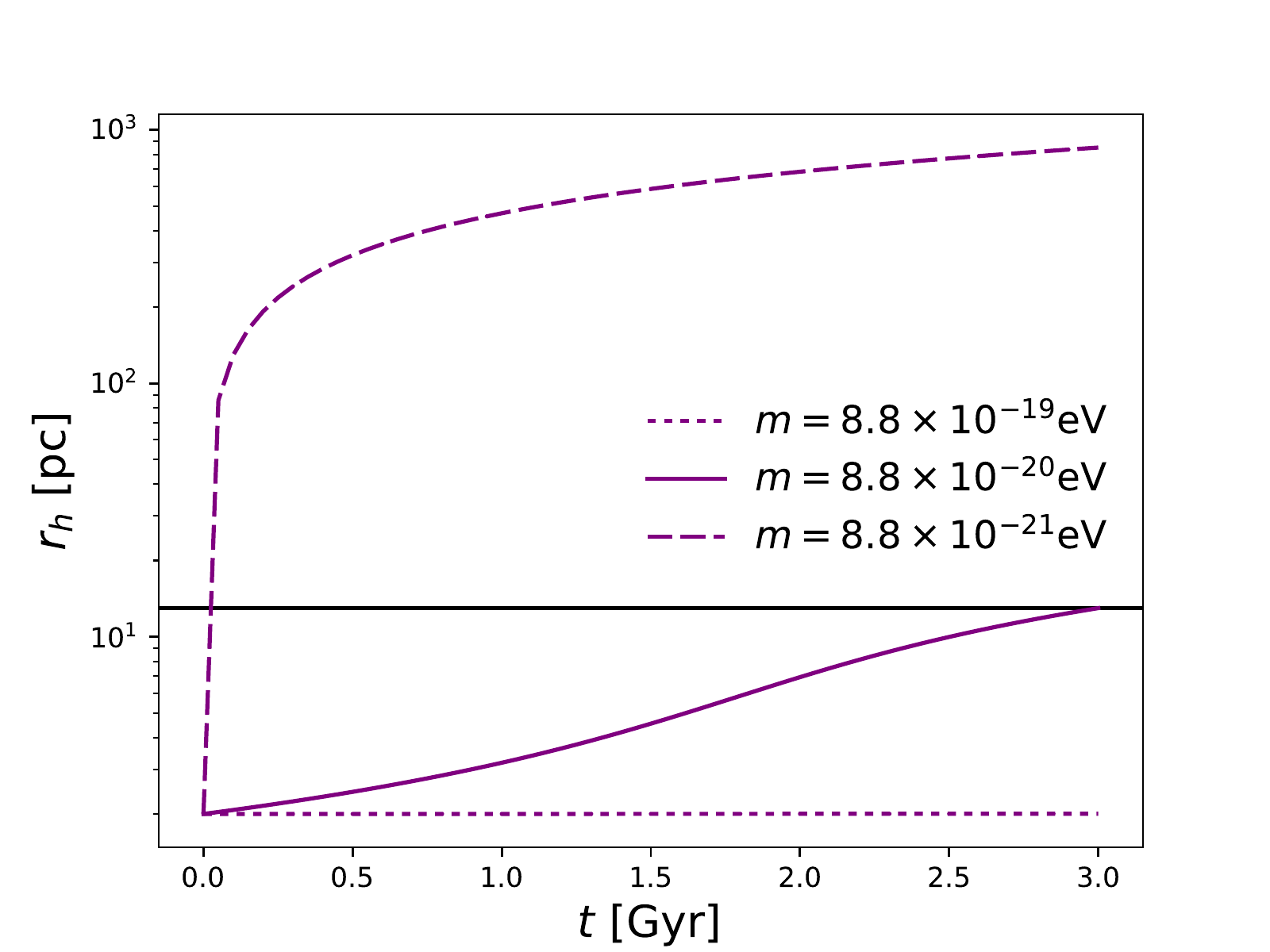}
	\caption{Expansion of the central star cluster of Eridanus II under the influence of fluctuations 
resulting from its embedding in a halo of FDM ultra-light axion dark matter of mass $m$. The horizontal 
line corresponds to the cluster's estimated half mass radius.}
	\label{fig:rhevol}
\end{figure}

 \cite{Marsh2018} applied the model of~\cite{EZFC} to study 
the effect of FDM fluctuations on the 
central cluster of the dwarf galaxy Eridanus II.  
As the cluster's embedding in a fluctuating  FDM halo can cause it to expand, the idea 
was to place constraints on the ultra light axion mass from the observed size of the cluster, in analogy 
with what~\cite{Brandt2016} obtained for MACHOS. 
As the present work presents an extension and refinement of the model of~\cite{EZFC}, 
with detailed application to the case 
of FDM haloes, it is of interest to revisit the case of Eridanus II in this 
context and examine its predictions. 

As in~\cite{Brandt2016} and~\cite{Marsh2018}, we consider the central cluster to expand in virial equilibrium, keeping the 
same form of the density profile, as its stars gain kinetic energy due to its embedding in the fluctuating medium. The equation for the temporal evolution 
of the cluster's half mass radius is then
\begin{equation}
\label{eq:app_MN18_dr}
\frac{dr_{h}}{dt} = \frac{D}{G} \left(\frac{\alpha M_\star}{r_h^2} +2\beta \rho_0 r_h\right)^{-1},
\end{equation}
where $D$ is the diffusion coefficient.
As in the aforementioned work we adopt $\alpha=0.4$ and $\beta=10$, which corresponds to a cored S\'ersic profile.
The star cluster mass is taken to be $M_\star = 2000 M_\odot$. 
Here we concentrate  on the case where the dark matter is composed solely of 
FDM. Furthermore, our model is strictly  valid  only outside the solitonic core, so the  factor $\mathcal{C F}$ in equation (16) 
of~\cite{Marsh2018} is unity (more on these issues below).    

In the context of the present work, the diffusion coefficient is given in terms of equation~(\ref{eq:max_disp}); that is  
\begin{equation}
\label{eq:app_MN18_Dv2_EZFCH}
D \left[ (\Delta v)^2 \right] = \frac{4 \sqrt{2} \pi G^2 \rho_0 m_{\rm eff}}{\sigma_{\rm eff}} \ln\Lambda \left[ \frac{{\rm erf}(X_{\rm eff})}{X_{\rm eff}} \right], 
\end{equation}
where $m_{\rm eff}$ is given by~(\ref{eq:meffM}),
$\sigma_{\rm eff}=\sigma/\sqrt{2}$ and $X_{\rm eff}= v_p/\sigma$. As 
in the previous appendix, we take the 
star cluster particle speed $v_p$ to correspond to the Keplerian velocity 
$v_p = 0.83~{\rm km/s} \sqrt{{\rm 13 pc}/R}$, and the dark matter 
velocity dispersion to be $\sigma = 6.9 ~{\rm km/s}$. Thus, ${\rm erf}(X_{\rm eff})/X_{\rm eff}$ is about 1.1. 
The Coulomb logarithm is given by Eq.~(\ref{eq:CouLEri}). Note that if the dark matter is not solely made of FDM, which instead
only constitutes a fraction $\mathcal{F}$ (as assumed in~\citealp{Brandt2016} 
and~\citealp{Marsh2018}) the diffusion coefficient  is effectively  
multiplied by a factor $\mathcal{F}^2$; as $\rho_0 \rightarrow \mathcal{F}\rho_0$ and 
 $m_{\rm eff} \rightarrow \mathcal{F} m_{\rm eff}$, since the latter also involves a factor $\rho_0$.

Fig.~\ref{fig:rhevol} shows the temporal evolution of the cluster's radius over an assumed age of 3 Gyr 
from an assumed initial radius of $r_h = 2~{\rm pc}$, which is the typical radius of stellar clusters in the Milky Way \citep[e.g.][]{Harris1996,Kharchenko2005}.
If the effect of FDM fluctuations is directed entirely at expanding the cluster, 
and if the current cluster radius  is $13 ~{\rm pc}$, then the results 
suggest an FDM axion mass $m \ge 8.8 \times 10^{-20} {\rm eV}$, which is essentially the same constraint
as that obtained by~\cite{Marsh2018} under the same assumptions regarding the FDM contribution
to the dark matter, the cluster's age and its initial and final sizes.  

However, as pointed out by~\cite{Marsh2018}, the cluster lies inside the solitonic core for masses $m  \la  10^{-20} {\rm eV}$. Strictly speaking neither the present formulation, 
or that of~\cite{EZFC} apply in this case. Indeed, in these contexts,  
the potential variations result from 
the existence of Gaussian random field of spatial fluctuations, transported into the time domain 
with mode velocity ${\bf v}$, rather than coherent core oscillations. Thus the 
constraints on the FDM axion mass appear limited in range.
For, although \cite{Marsh2018} use a factor $\mathcal{C}$ (taken to be 0.3 inside the core)
to take into account the effect of attenuated amplitudes of fluctuations due to core oscillations, 
and thus extend the constraint to smaller FDM masses  for which the cluster lies inside the core, 
as Fig.~\ref{fig:Corr} (upper panel) shows, the fluctuations close to the core also have much longer correlation times. 
This may render the random Gaussian assumption for the fluctuations, and the associated stochastic dynamics with a diffusion limit, more
difficult to justify  unless the timescales  considered are sufficiently long.
As mentioned in Section~\ref{sec:typical_dens}, incorporating the core oscillations would in addition require 
modification of the model to take into account the associated limited frequency range characterizing the oscillations. 

These complications may not entirely eliminate the effect of core oscillations, and  
fluctuations from FDM granules outside the core can still 
affect the evolution of a cluster embedded inside it, and thus lead to strong constraints on $m$. 
However another caveat hinders direct extensions of the exclusion limits obtained above to small masses.  
Namely, for masses  greater than $m = 1.6 \times 10^{-19} {\rm eV}$, the minimal 
wavelength associated with the size of the FDM quasi-particle granules is 
larger than the initial estimated characteristic cluster size: $\lambda_{\rm min} \le 2 {\rm pc}$.  
For $m \approx 10^{-20} {\rm eV}$ it is an order of magnitude larger. 
It is therefore unclear whether it is appropriate to consider that the associated fluctuations
(or core oscillations) 
affect the internal structure of the cluster, as assumed above,  
rather than  the  cluster as a whole. 
If the latter situation is assumed, then for $m_{\rm eff} \gg M_\star$ one may expect energy equipartition 
between FDM quasi-particles and the cluster to result in significant motion of its centre of mass. 

For example, 
for $m = 10^{-20} {\rm eV}$, $m_{\rm eff} = 1.8  \times 10^4~M_\odot \gg M_\star$, which 
is an order of magnitude larger than the total cluster mass. The cluster should then gain 
random motion as a result of its embedding in the FDM 'heat bath'. 
Indeed, in that case, naive application 
of Eq.~(\ref{eq:veldispt}) leads one to deduce that the cluster gains a random velocity of order $\langle (\Delta v_c)^2 \rangle^{1/2}
= 24 ~{\rm km/s}$, which should displace it well outside the centre of the galaxy.  

Such arguments, based on the 
cluster's displacement form the centre of the galaxy, may thus in principle 
rule smaller out FDM axion masses, replacing the ones based on the cluster's expansion.  
However, detailed  examination of such issues, 
including the more complex intermediate mass case, are beyond the  scope of this work.


\bsp	
\label{lastpage}
\end{document}